\documentclass[groupedaddress,twocolumn,10pt,prl,aps,showpacs,longbibliography,floatfix]{revtex4-2}

\usepackage[dvipsnames]{xcolor}

\usepackage{lipsum}
\usepackage{physics}
\usepackage{graphicx}
\usepackage{dcolumn}
\usepackage{bm}
\usepackage[breaklinks=true,colorlinks,citecolor=blue,linkcolor=blue,urlcolor=blue]{hyperref}
\usepackage{titlesec}
\usepackage[caption=false]{subfig}

\renewcommand{\paragraph}[1]{\textit{#1}---.}

\usepackage[capitalise]{cleveref}

\usepackage{titlesec}
\titleformat{\section}[runin]
{\normalfont\itshape}{\thesection}{0em}{}[.]

\usepackage{mathtools}
\usepackage{amsmath}
\allowdisplaybreaks
\usepackage{ bbold }
\usepackage{booktabs}
\usepackage{amssymb}
\usepackage[normalem]{ulem}
\usepackage{amsthm}
\usepackage{xfrac}
\usepackage{dsfont}
\usepackage{siunitx}
\usepackage{physics}
\usepackage{graphicx}
\usepackage{hyperref}
\usepackage[capitalise]{cleveref}
\usepackage{dcolumn}
\usepackage{bm}
\usepackage{layouts}
\usepackage{comment}

\newcommand{\rme}{{\rm e}}
\newcommand{\rmg}{{\rm g}}
\newcommand{\rmd}{{\rm d}}
\newcommand{\rmc}{{\rm c}}
\newcommand{\rma}{{\rm a}}
\newcommand{\rmb}{{\rm b}}
\newcommand{\rmi}{{\rm i}}
\newcommand{\rmf}{{\rm f}}
\newcommand{\rms}{{\rm s}}
\newcommand{\rmt}{{\rm t}}
\newcommand{\rmT}{{\rm T}}
\newcommand{\bfr}{{\bf r}}
\newcommand{\bfk}{{\bf k}}
\newcommand{\bfn}{{\bf n}}
\newcommand{\bfd}{{\bf d}}

\newcounter{suppinfonum}

\newcommand{\suppinfofootnote}{%
  \footnote{For more details, see the Supplementary Information.}%
  \setcounter{suppinfonum}{\value{footnote}}%
}

\newcommand{\suppinfo}{\footnotemark[\value{suppinfonum}]}

\begin{document}

\author{Filippo Ferrari}
\email{filippo.ferrari@epfl.ch}
\author{Francesca Orsi}
\author{Ekaterina Fedotova}
\author{Óscar Rios Alves}
\author{Michał Zdziennicki}
\author{Jean-Philippe Brantut}
\author{Vincenzo Savona}
\email{vincenzo.savona@epfl.ch}
\affiliation{Institute of Physics and Center for Quantum Science and Engineering,\\ \'{E}cole Polytechnique F\'{e}d\'{e}rale de Lausanne (EPFL), Lausanne, Switzerland}

\title{Controlling many-body quantum chaos in a dissipative optical cavity}

\date{\today}
             
\begin{abstract}
Cavity quantum electrodynamics (QED) with ultracold fermions provides a promising platform for realizing many-body quantum chaos through disordered, photon-mediated long-range interactions. 
Such setups are inherently open and are therefore subject to dissipation arising from cavity photon loss and atomic spontaneous emission.
In this article, we study the driven-dissipative dynamics of a typical cavity QED setting including controllable disorder and long-range interactions. We find that the two dissipation sources have qualitatively different structures. Cavity loss reduces to a single dephasing channel, whereas spontaneous emission in the experimentally relevant regime generates a collection of nonlocal dephasing channels. 
Cavity-induced dephasing preserves signatures distinguishing integrable from chaotic Hamiltonian dynamics in observables that depend linearly on the density matrix, while spontaneous emission suppresses these signatures.
By contrast, quantities that probe the structure of the many-body state, such as the entanglement entropy, are strongly affected by both dissipation mechanisms. 
Assuming experimentally realistic parameters, we derive quantitative constraints for the observation and control of many-body quantum chaos in cavity-QED platforms.
\end{abstract}

\maketitle

\paragraph{Introduction}
Cavity quantum electrodynamics (QED) with ultracold atoms has emerged as a versatile platform for analog quantum simulation~\cite{ritsch_cold_2013, georgescu_quantum_2014, mivehvar_cavity_2021, altman_quantum_2021}. 
In the past decades, significant experimental milestones have been achieved by the community, including the observation of density-wave ordering in bosonic~\cite{baumann_dicke_2010, baumann_exploring_2011, mottl_roton-type_2012, landig_quantum_2016, leonard_supersolid_2017} and fermionic~\cite{zhang_observation_2021, helson_density-wave_2023, zwettler_cavity-mediated_2025, buhler_microscopy_2026} gases, the realization of spin-glass~\cite{guo_sign-changing_2019, kroeze_directly_2025, marsh_multimode_2025} and time-crystalline~\cite{kesler_observation_2021, kongkhambut_observation_2022} phases, advances in quantum metrology~\cite{leroux_implementation_2010, chen_conditional_2011, hosten_measurement_2016, huang_observing_2023, robinson_direct_2024}, and the characterization of nonequilibrium quantum phase transitions~\cite{brennecke_real-time_2013, klinder_dynamical_2015, dogra_dissipation-induced_2019, muniz_exploring_2020, ferri_emerging_2021, young_observing_2024, song_dissipation-induced_2025}. 
Crucially, these phenomena arise from cavity-mediated long-range interactions, which render the system effectively collective and, in many regimes, well described by mean-field or semiclassical theories~\cite{domokos_collective_2002, nagy_dicke-model_2010, ritsch_cold_2013, mivehvar_cavity_2021, dogra_phase_2016, jager_mean-field_2016, himbert_mean-field_2019, defenu_long-range_2023}.

\begin{figure}[t]
\centering
\includegraphics[width=0.45\textwidth]{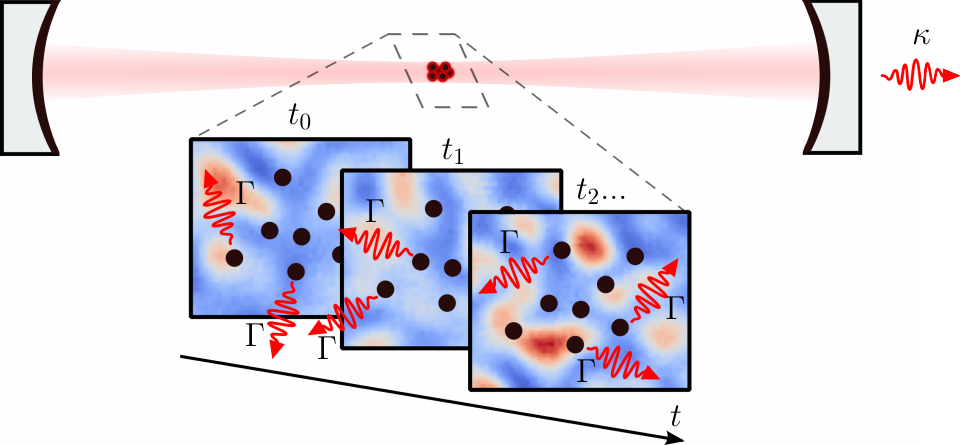}\vspace{0.05em}
\caption{
Sketch of the experimental setup: a high-finesse optical cavity is coupled to a cloud of driven ultracold fermionic atoms.
The light-matter interaction is disordered by an engineered, time-dependent speckle potential projected on the atomic ensemble, thus realizing a quantum chaotic many-body system.
Two sources of decoherence disturb the unitary quantum dynamics: the cavity dissipation at a rate $\kappa$ and the atomic spontaneous emission at a rate $\Gamma$.
}
\label{fig:sketch}
\end{figure}

Leveraging the cavity-mediated interaction to realize states with strong quantum correlations is an outstanding experimental frontier. Indeed, cavity QED platforms are inherently open quantum systems, subject to two primary sources of decoherence: photon leakage from the cavity and incoherent photon scattering by atoms (see Fig.~\ref{fig:sketch} for a sketch). The extent to which cavity-QED quantum simulators can faithfully realize strongly correlated and dynamically complex many-body phenomena under realistic experimental conditions is therefore a central question.
In this work, we address it in the context of many-body quantum chaos, which fundamentally relies on the buildup of quantum correlations, entanglement, and operator spreading beyond effective collective descriptions.

Quantum chaos is a universal behavior expected to emerge in generic interacting and isolated quantum many-body systems \cite{dalessio_quantum_2016, haake_quantum_2018}.
A paradigmatic class of chaotic quantum systems is the Sachdev-Ye-Kitaev (SYK) type of models \cite{sachdev_gapless_1993, kitaev_simple_2015, maldacena_bound_2016, maldacena_remarks_2016, rosenhaus_introduction_2019}, realized by fermionic modes subject to random, long-range interactions. 
The interest around random fermionic models stems from the broad range of fundamental phenomena they provide access to, ranging from strongly correlated phases of matter \cite{sachdev_bekenstein-hawking_2015, davison_thermoelectric_2017, song_strongly_2017, esterlis_cooper_2019, wang_solvable_2020, chowdhury_sachdev-ye-kitaev_2022, patel_universal_2023} and quantum thermalization \cite{sonner_eigenstate_2017}, to holography and black-hole physics \cite{hayden_black_2007, maldacena_conformal_2016, cotler_black_2017, kitaev_soft_2018}.
Cavity QED with ultracold fermions \cite{ritsch_cold_2013, mivehvar_cavity_2021} provides a promising platform for the experimental realization and control of such systems, 
thanks to the simultaneous presence in a single setup of i) natively fermionic particles, ii) long-range interactions \cite{mivehvar_cavity_2021}, iii) controllable disorder in the couplings \cite{sauerwein_engineering_2023, orsi_cavity_2024}.
A number of theoretical proposals argued that cavity QED architectures can engineer random chaotic fermionic systems, including the SYK model \cite{uhrich_cavity_2023, baumgartner_quantum_2024, baumgartner_quantum_2025, solis_single-particle_2026}.

In this paper, we study the driven-dissipative dynamics of a cavity-fermion setup realizing, in the absence of dissipation, many-body quantum chaos mediated by virtual photons. 
Considering realistic parameters~\cite{bolognini_design_2025}, we find that dissipation cannot be neglected in any experimentally accessible regime. 
Our central finding is that the two decoherence sources have fundamentally different structures: cavity dissipation reduces to a single dephasing channel, while spontaneous emission typically generates a large collection of non-local dephasing channels. The former preserves thermalization fingerprints in observables that are linear in the density matrix, allowing an experimental distinction between integrable and chaotic dynamics; the latter erases any such distinction. Quantities relying on the wave function structure, such as entanglement, are degraded by both sources. 

Our results can be extended to cavity QED simulators operating at realistic parameters and aiming at building up long-range quantum coherence and provide concrete constraints for the design of experiments in this class of ultracold atomic platforms.

\paragraph{System and protocols}\label{sec:model}
We consider a 2D gas of ultracold fermions, trapped within a single-mode optical cavity, described by the bosonic annihilation (creation) operator $\hat{a}$ ($\hat{a}^\dagger$).
The motional eigenstates of the harmonic trap provide modes or orbitals, with $\hat{c}_j$ ($\hat{c}_j^\dagger$) annihilating (creating) a fermion in the $j$-th mode.
In the presence of a laser pump far detuned from the atomic transition, the Hamiltonian describing the system reads
\begin{equation}\label{eq:hamiltonian}
    \hat{H} = \Delta_{\rmc\rmd}\hat{a}^\dagger\hat{a} 
    + \frac{\Omega\Omega_\rmd}{2\Delta_{\rmd\rma}}\sum_{jk}g_{jk}\,(\hat{a}^\dagger + \hat{a})\,\hat{c}_j^\dagger\hat{c}_k.
\end{equation}
In Eq.~\eqref{eq:hamiltonian}, $\Delta_{\rmc\rmd} = \omega_\rmc - \omega_\rmd$ is the cavity-to-drive detuning, $\Delta_{\rma\rmd} = \omega_\rma - \omega_\rmd$ is the atom-to-drive detuning, with $\omega_\rmc$, $\omega_\rma$ and $\omega_\rmd$ the cavity, atomic and drive frequencies, respectively.
$\Omega$ is the atom-cavity Rabi frequency and $\Omega_\rmd$ is the drive amplitude.
Finally, $g_{jk}$ are disordered, all-to-all couplings, resulting from the orbitals hosting the atoms and a speckle potential projected on them~\cite{goodman_fundamental_1976}, which induces \textit{controllable randomness}~\suppinfofootnote.

From now on, we assume $\Delta_{\rmc\rmd}$ to be large, i.e., no physical photons are present and virtual photons mediate long-range interactions among the fermions. The effective, fermionic-only Hamiltonian reads
\begin{equation}\label{eq:hamiltonian_effective}
    \hat{H}_{\rm eff} = -\frac{1}{\Delta_{\rmc\rmd}}\frac{\Omega^2\Omega_\rmd^2}{4\Delta_{\rmd\rma}^2}\left[\sum_{jk}g_{jk}\hat{c}_j^\dagger\hat{c}_k\right]^2.
\end{equation}
As already noticed in Refs.~\cite{kim_low-rank_2020, uhrich_cavity_2023, baumgartner_quantum_2024, baumgartner_quantum_2025}, the above Hamiltonian does not show quantum chaos because of the low-rank structure of the two-body interaction tensor~\footnote{Namely, in Eq.~\eqref{eq:hamiltonian_effective} the two-body term $\hat{c}_j^\dagger\hat{c}_\ell^\dagger\hat{c}_k\hat{c}_m$ is mediated by $g_{jk}g_{\ell m}$, which is not the rank-4 tensor of SYK-type models $g_{jk\ell m}$~\cite{chowdhury_sachdev-ye-kitaev_2022}. This makes the model nonchaotic.}.

\begin{figure*}[t]
\centering
\includegraphics[width=\textwidth]{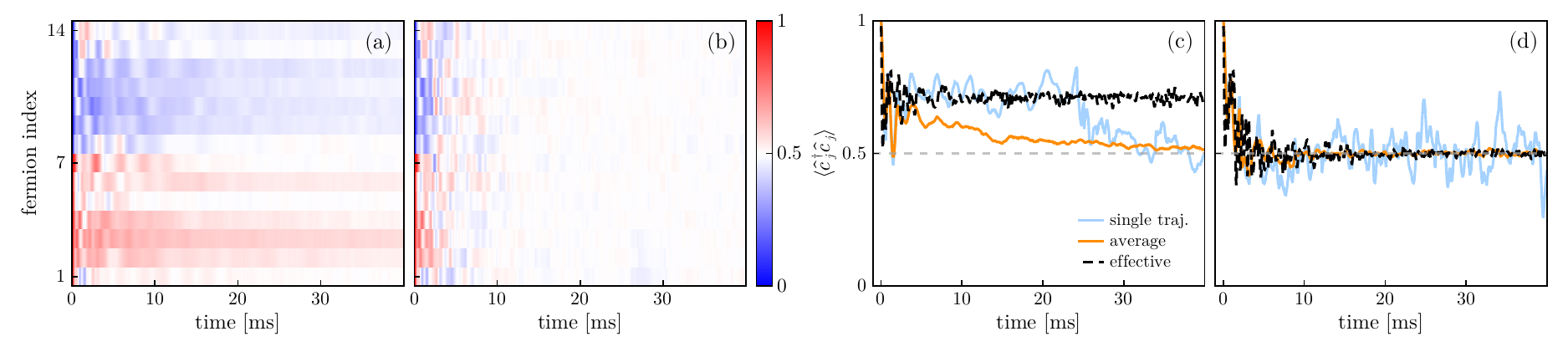}\vspace{0.05em}
\caption{
(a-b) Averaged dynamics of $\langle\hat{c}_j^\dagger\hat{c}_j\rangle$ for $N=14$ fermionic modes at half filling.
We consider the cavity-fermion system described in Eq.~\eqref{eq:hamiltonian} with (a) static disorder and (b) dynamical disorder [the f-RQC protocol in Eq.~\eqref{eq:f_rqc}], accounting for cavity dissipation at rate $\kappa$ and spontaneous emission at rate $\Gamma$.
(c-d) Comparison between the dissipative quantum trajectory averaged dynamics (orange curves), the single-trajectory averaged dynamics (light blue curves) and the effective dynamics generated by Eq.~\eqref{eq:hamiltonian_effective} with $\kappa,\,\Gamma=0$ (black-dahsed curves) for (c) static disorder and (d) the f-RQC protocol, and for $j=4$.
As initial state we choose $\ket{\Psi(0)} = \ket{1\,...\,1\,0\,...\,0}$.
Parameters are fixed to 
$\Omega/2\pi=4.85\,\textrm{MHz}$,
$\kappa/2\pi=200\,\textrm{kHz}$,
$\Gamma/2\pi=5.9\,\textrm{MHz}$ (hence $\mathcal{C}=20$),
$\Delta_{\rmc\rmd}/2\pi = 100\,\textrm{kHz}$,
$\Omega_\rmd/2\pi = 12\,\textrm{MHz}$,
$\Delta_{\rmd\rma}/2\pi=3\,\textrm{GHz}$.
The number of disorder patterns used for the f-RQC is fixed to $n=2N+1$.
The number of stochastic quantum trajectories is fixed to $N_{\rm traj}=100$.
}
\label{fig:small_detuning}
\end{figure*}

To obtain a quantum chaotic model, Refs.~\cite{baumgartner_quantum_2024, baumgartner_quantum_2025} proposed to rapidly change the disorder pattern over time. 
In this paper, we study a simplified protocol that we dub a fermionic random quantum circuit (f-RQC), leading to a unitary time evolution described by
\begin{equation}\label{eq:f_rqc}
    \hat{U}(t) = \prod_{\alpha=1}^n \rme^{-\rmi\hat{H}^{(\alpha)}_{\rm eff} \Delta t},
\end{equation}
with $n\sim\mathcal{O}(N)$, and $\alpha$ running over $n$ different disorder realizations. 
Eq.~\eqref{eq:f_rqc} generates many-body quantum chaos in the dissipationless fermionic gas~\suppinfo.

\paragraph{Dissipation and relevant time scales}
There are two sources of dissipation: i) the single-photon losses from the cavity at the rate $\kappa$, and ii) atomic spontaneous emission at rate $\Gamma$. Because the atomic excited states are adiabatically eliminated, the latter is a Rayleigh scattering process in which a ground-state atom elastically scatters a drive photon into free space, recoiling by $k_0 = 2\pi/\lambda_\rmc$ with $\lambda_\rmc$ the cavity wavelength.
The fundamental parameter of cavity QED experiments is the single-atom cooperativity $\mathcal{C}=\Omega^2/\kappa\Gamma$, i.e., the ratio between coherent emission of the atom in the cavity mode, with respect to its incoherent emission over the full solid angle.
Throughout the paper, we fix $\mathcal{C}=20$~\cite{bolognini_design_2025}.
Furthermore, here we assume we work with $^6$Li atoms: this fixes $\lambda_\rmc$ and $\Gamma$, but the conclusions we derive in the main text are independent of this choice.
In the End Matter we discuss more in detail how the atomic species choice can influence our treatment via additional spurious terms appearing in Eq.~\eqref{eq:hamiltonian} and we propose some schemes to circumvent unwanted effects.

There are three relevant time scales in the dispersive regime~\suppinfo: the coherent time scale set by $\mathcal{E}$, and the two dissipative time scales set by $\kappa_{\rm eff}$ and $\Gamma_{\rm eff}$,
\begin{equation}\label{eqs:time_scales}
    \mathcal{E} \sim \frac{\Omega^2\Omega_\rmd^2}{|\Delta_{\rmc\rmd}|\Delta_{\rmd\rma}^2},\quad
    \kappa_{\rm eff} \sim \kappa\frac{\Omega^2\Omega_\rmd^2}{\Delta_{\rmc\rmd}^2\Delta_{\rmd\rma}^2}\quad 
    \Gamma_{\rm eff} \sim \Gamma\frac{\Omega_\rmd^2}{\Delta_{\rmd\rma}^2}.
\end{equation}
An important point of our discussion is that $\kappa_{\rm eff}$ and $\Gamma_{\rm eff}$ must be studied separately against $\mathcal{E}$.
This yields $\mathcal{E}/\kappa_{\rm eff} = |\Delta_{\rmc\rmd}|/\kappa$ and  $\mathcal{E}/\Gamma_{\rm eff} = \Omega^2/|\Delta_{\rmc\rmd}|\Gamma$.
We recognize $|\Delta_{\rmc\rmd}|$ as the control parameter between a regime where cavity dissipation dominates and a regime where spontaneous emission dominates, assuming fixed $\Omega_\rmd$ and $\Delta_{\rma\rmd}$.
Even for a cooperativity $\mathcal{C}\sim20$ the coherent dynamics at the single-atom level is not immune from dissipation regardless of the choice of $|\Delta_{\rm cd}|$.

\paragraph{Quantum many-body dynamics}
To simulate the dissipative quantum many-body dynamics of $N$ fermionic modes at half filling coupled to a cavity, we resort to Monte Carlo stochastic quantum trajectories \cite{dalibard_wave-function_1992, molmer_monte_1993, daley_quantum_2014}.
Cavity dissipation is described by the Lindblad jump operator $\hat{L}_\rmc = \sqrt{\kappa}\hat{a}$.
Atomic spontaneous emission is modeled via a collection of non-local dephasing jump operators in the form $\hat{L}^{(r)}_\rma\propto\sqrt{\Gamma_{\rm eff}}\sum_{jk}M_{jk}^{(r)}\hat{c}_j^\dagger\hat{c}_k$~\suppinfo.
This description encodes the recoil imparted by each scattering event, which drives heating and incoherent transitions between orbitals.
The fundamental parameter shaping the features of spontaneous emission is the Lamb-Dicke parameter $\eta=x_0k_0 = 2\pi x_0/\lambda_\rmc$ where $x_0$ is typical length of the trap.
Intuitively, $\eta$ measures the transition amplitude between neighboring orbital upon photon scattering.
Realistically, for $^6$Li atoms in a typical tweezer trap, $\lambda_\rmc=671\,\textrm{nm}$ and $x_0=100\,\textrm{nm}$, yielding $\eta\simeq0.94$.
For such values of $\eta$, recoil processes matter in the fermionic dynamics~\footnote{Throughout the paper, we always work within a finite set of orbitals, which encodes the sites $j$ of the random fermionic model in Eq.~\eqref{eq:hamiltonian_effective}. A detrimental effect of heating is atom loss, i.e., the ejection of atoms outside the trap.
In the End Matter we argue that this effect can be eliminated with a careful engineering of the trap, leaving the intra-manifold heating as the dominant process caused by atomic spontaneous emission.}.
As we show below, spontaneous emission at $\eta\sim{O}(1)$ is a major obstacle for the control of many-body quantum chaos in optical cavities operating at $\mathcal{C}$ of a few tens.
In the End Matter, we consider the comparison with $\eta\ll1$ (the so-called Lamb-Dicke regime).

We consider the initial state $\ket{\Psi(0)} = \ket{1\,...\,1\,0\,...\,0}$ and we track the time evolution of the fermionic population $\langle\hat{c}_j^\dagger\hat{c}_j\rangle$ along single quantum trajectories.
Addressing single atoms is feasible in experiments with ultracold fermions~\cite{holten_observation_2021, holten_observation_2022}.
The observable $\langle\hat{c}_j^\dagger\hat{c}_j\rangle$ encodes fingerprints of integrability and chaos in the quantum dynamics.
Integrability implies persistent oscillations or a clear memory of the initial state structure, while chaos yields $\langle\hat{c}_j^\dagger\hat{c}_j\rangle\sim 0.5$ for fermions at half filling~\cite{dalessio_quantum_2016}.
In addition to $\mathcal{C}$, in all the simulations we also fix $\kappa/2\pi=200\,\textrm{kHz}$.

\begin{figure*}[t]
\centering
\includegraphics[width=\textwidth]{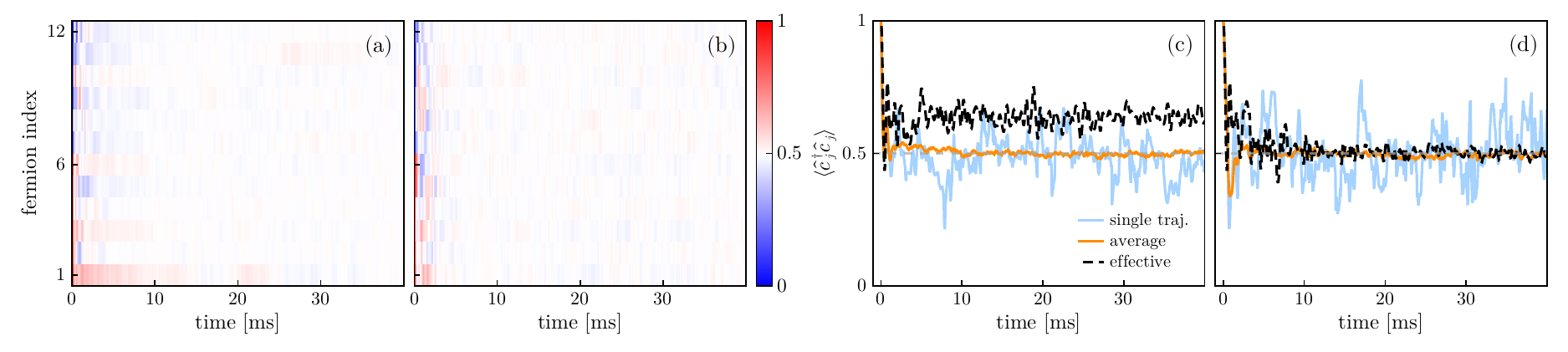}\vspace{0.05em}
\caption{
Same as in Fig.~\ref{fig:small_detuning} but for $\Delta_{\rmc\rmd}/2\pi=1\,\textrm{MHz}$.
The drive amplitude has been increased to $\Omega_\rmd/2\pi=38\,\textrm{MHz}$ to match the same $\mathcal{E}$ in the simulations reported in Fig.~\ref{fig:small_detuning}.
The number of fermionic modes is $N=12$.
Other parameters as in Fig.~\ref{fig:small_detuning}.
}
\label{fig:large_detuning}
\end{figure*}

First, we consider $\Delta_{\rmc\rmd}=\kappa/2$, corresponding to cavity dissipation dominating over spontaneous emission and comparable to the coherent energy scale [$\mathcal{E}/\kappa_{\rm eff}\sim\mathcal{O}(1)$].
In Figs.~\ref{fig:small_detuning} (a) and (b) we plot the dynamics of $\langle\hat{c}_j^\dagger\hat{c}_j\rangle$ for the static disorder case and the f-RQC protocol respectively, averaged over $N_{\rm traj}=100$ stochastic quantum trajectories and for $N=14$ fermionic modes.
In the case of static disorder, we observe that the dynamics of $\langle\hat{c}_j^\dagger\hat{c}_j\rangle$ retains a memory about the structure of the initial state, a signature of integrability.
In the case of the f-RQC, instead, $\langle\hat{c}_j^\dagger\hat{c}_j\rangle$ uniformly approaches $\langle\hat{c}_j^\dagger\hat{c}_j\rangle\sim 0.5$.
In this case, no reminiscence about the initial state is visible and the fermionic density converges to the thermal value for half filling~\cite{sonner_eigenstate_2017}.
Next, we compare in Figs.~\ref{fig:small_detuning} (c) and (d) the dissipative dynamics with the unitary dynamics generated by Eq.~\eqref{eq:hamiltonian_effective} for $j=4$.
While the effective (black-dashed curve) and averaged dynamics (orange curve) are quantitatively different, we observe that the behavior of $\langle\hat{c}_j^\dagger\hat{c}_j\rangle$ is qualitatively the same: in both the cases for static disorder the fermionic density does not converge to the thermal value, while it does for the f-RQC protocol~\footnote{For long times, we observe a decay of $\langle\hat{c}_j^\dagger\hat{c}_j\rangle$ towards the thermal value. As we show in the Supplementary Information, this is entirely due to the onset of spontaneous emission-induced dephasing an not to the cavity-induced dephasing.}. 
We then investigate the behavior of single stochastic trajectories in both cases [light blue curves in Figs.~\ref{fig:small_detuning} (c) and (d)].
The single trajectory differs from the averaged dynamics, signaling that the system density matrix starts to be mixed due to dissipation.
Looking at the f-RQC dynamics, we observe how $\langle\hat{c}_j^\dagger\hat{c}_j\rangle$ over a single trajectory does not converge to the expected thermal value, but exhibits large oscillations around it.
The same conclusion holds for the other $j$ (not shown).
This implies that fingerprints of many-body quantum chaos are lifted at the level of single quantum trajectories, but not for the averaged dynamics.

We now consider $\Delta_{\rmc\rmd}=5\kappa$, corresponding to spontaneous emission dominating over cavity dissipation and comparable to the coherent energy scale [$\mathcal{E}/\Gamma_{\rm eff}\sim\mathcal{O}(1)$].
The action of the high-rank dissipator is made manifest in Figs.~\ref{fig:large_detuning} (a) (static disorder) and (b) (f-RQC protocol): any difference between the underlying integrable or chaotic Hamiltonian behavior is canceled and the averaged fermion density $\langle\hat{c}_j^\dagger\hat{c}_j\rangle$ uniformly approach the thermal value $\langle\hat{c}_j^\dagger\hat{c}_j\rangle\sim 0.5$.
In other words, the cavity QED simulator is no longer able to control the unitary dynamics generated by the coherent fermionic interaction.
This effect is even more evident in Figs.~\ref{fig:large_detuning} (c) and (d), where we compare effective and dissipative dynamics for the static disorder and the f-RQC protocols, respectively.
In both cases, the averaged $\langle\hat{c}_j^\dagger\hat{c}_j\rangle$ converges to the thermal value while the single-trajectory $\langle\hat{c}_j^\dagger\hat{c}_j\rangle$ fluctuates around it.
We conclude that the large-detuning regime, which seems naively better as it suppresses cavity dissipation, is the most exposed to the detrimental action of spontaneous emission on the quantum coherences.

We propose an intuitive argument for which cavity dissipation and spontaneous emission have so different effects on the coherent dynamics of Eq.~\eqref{eq:hamiltonian_effective}.
In the dispersive regime, $\hat{L}_\rmc=\sqrt{\kappa}\hat{a}$ amounts to an effective, non-local dephasing on the fermions described by the Lindblad dissipator \cite{breuer_theory_2007}
\begin{equation}\label{eq:cavity_effective_dissipator}
    \mathcal{D}\hat{\rho}\sim\kappa_{\rm eff}\sum_{jk\ell m}H_{jk}H_{\ell m}\hat{c}_j^\dagger\hat{c}_k\,\hat{\rho}\,\hat{c}_\ell^\dagger\hat{c}_m,
\end{equation}
where $H_{jk}$ are disordered couplings correlated with the $g_{jk}$~\suppinfo.
We recognize in Eq.~\eqref{eq:cavity_effective_dissipator} a sparse disorder structure similar to the one arising in the effective, non-chaotic model in Eq.~\eqref{eq:hamiltonian_effective}. 
This is equivalent to a \textit{single} Lindblad jump operator $\hat{L}_\rma^{({\rm eff})}=\sqrt{\kappa_{\rm eff}}\sum_{jk}H_{jk}\hat{c}_j^\dagger\hat{c}_k$.
Away from Lamb-Dicke regime, spontaneous emission is instead associated to the dissipator~\suppinfo
\begin{equation}\label{eq:spontaneous_emission_effective_dissipator}
    \mathcal{D}\hat{\rho} \sim \Gamma_{\rm eff} \sum_{jk\ell m}K_{jk\ell m}\hat{c}_j^\dagger\hat{c}_k\,\hat{\rho}\,\hat{c}_\ell^\dagger\hat{c}_m,
\end{equation}
Crucially, $K_{jk\ell m}$ has a dense disorder structure similar to the random interaction in the chaotic SYK model~\cite{chowdhury_sachdev-ye-kitaev_2022}. 
This corresponds to an extensive collection of non-local dephasing jump operators.
While $\hat{L}_\rma^{({\rm eff})}$ changes the averaged dynamics only quantitatively, the spontaneous emission erases any signature of integrability as it results from Fig.~\ref{fig:large_detuning} (a).
Our numerical data indicate that the different microscopic decoherence channels project into qualitatively different effective dissipators after cavity elimination, and this distinction determines whether experimentally accessible chaos signatures survive or not.

\begin{figure*}[t]
\centering
\includegraphics[width=\textwidth]{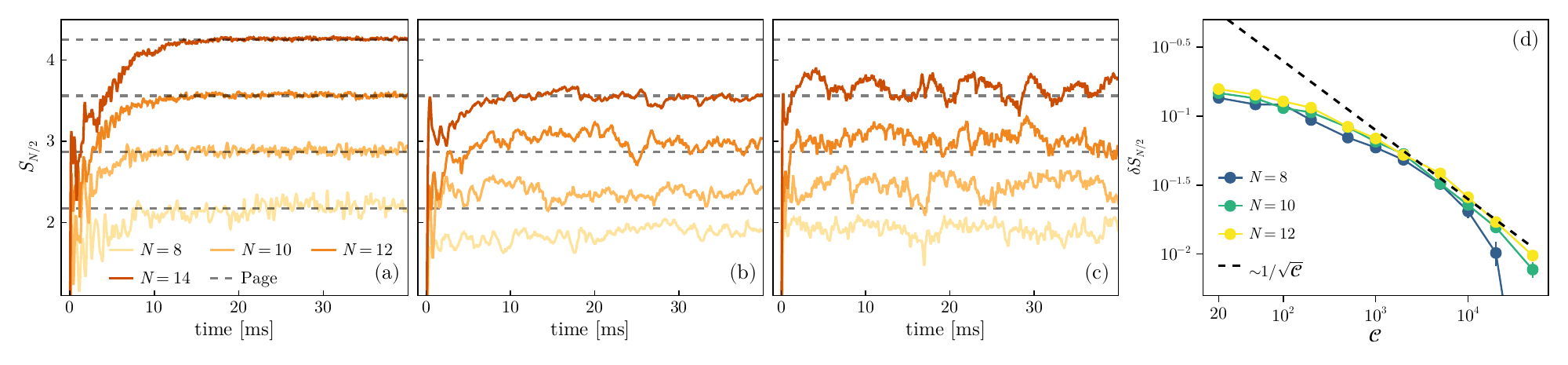}\vspace{0.05em}
\caption{
Entanglement entropy dynamics for (a) the effective model in Eq.~\eqref{eq:hamiltonian_effective}, (b-c) the full model in Eq.~\eqref{eq:hamiltonian} with nonzero $\kappa,\,\Gamma$, for $\Delta_{\rmc\rmd}=\kappa/2$ and $\Delta_{\rmc\rmd}=5\kappa$ respectively.
We consider the f-RQC protocol with $n=2N+1$.
The gray-dashed line represent to the Page prediction for a $\mathbb{U}(1)$ symmetric system at half filling.
Results are reported for $N=8,\,10,\,12,\,14$ fermionic modes at half filling (from light to dark orange) and are averaged over $N_{\rm traj}=100$ trajectories [except $N=14$ in panel (c), for which $N_{\rm traj}=20$].
(d) Relative distance from Page entropy $\delta S_{N/2} = |S_{N/2}-S_{\rm Page}|/S_{\rm Page}$ as a function of $\mathcal{C}$ for $\Delta_{\rm cd}=2\kappa$ ($\Omega_\rmd/2\pi=25\,\textrm{MHz}$) and different $N$.
Data are averaged over $N_{\rm traj}=100$ trajectories and 5 unitaries~\eqref{eq:f_rqc}.
The black-dashed line represents $\sim1/\sqrt{\mathcal{C}}$.
Other parameters as in Fig.~\ref{fig:small_detuning}.
}
\label{fig:entanglement_entropy}
\end{figure*}

\paragraph{Quantum many-body entanglement}
In this paragraph we study the entanglement dynamics. 
From the experimental point of view, accessing entanglement in ultracold atomic platforms is in principle possible~\cite{islam_measuring_2015, kaufman_quantum_2016}, but it remains very challenging especially in the large-$N$ limit.
From a theoretical perspective, the fermionic entanglement entropy provides a useful diagnostic of quantum chaos and thermalization~\cite{dalessio_quantum_2016}, while also characterizing the structure of the many-body wave function through the degree of mixedness of a subsystem. Here, we compute the entanglement entropy of a subsystem comprising half of the fermionic degrees of freedom, $\hat{\rho}_\rmf = \operatorname{Tr}_\rmc[\ketbra{\Psi(t)}]$, $S_{N/2} = -\operatorname{Tr}\left[\operatorname{Tr}_{N/2}(\hat{\rho}_\rmf)\,\log\operatorname{Tr}_{N/2}(\hat{\rho}_\rmf)\right]$. 
For maximally chaotic closed quantum systems, the dynamics generates states whose entanglement entropy satisfies Page's prediction \cite{page_average_1993, sen_average_1996}.
In the presence of a  $\mathbb{U}(1)$ symmetry and for $N/2$ fermions traced out, the Page entropy reads $S_{\rm Page} = \frac{N}{2}\log(2) - \frac{1}{2}\log(2) - \frac{1}{4}$~\cite{yauk_typical_2024}. 

In Fig.~\ref{fig:entanglement_entropy} we plot the dynamics of $S_{N/2}$ for the f-RQC protocol corresponding to (a) the effective model in Eq.~\eqref{eq:hamiltonian_effective} and the full model in Eq.~\eqref{eq:hamiltonian} including dissipation for (b) $\Delta_{\rmc\rmd}=\kappa/2$ and (c) $\Delta_{\rmc\rmd}=5\kappa$.
We observe how the entanglement entropy saturates the theoretical bound for the effective model with $N=8,\,10,\,12,\,14$ fermionic modes, confirming that the f-RQC protocol generates quantum chaos.
When dissipation is included, the half-system entanglement entropy $S_{N/2}$ remains below $S_{\rm Page}$ for both values of $\Delta_{\rm cd}$. 
This behavior can be understood within the framework of monitored many-body quantum dynamics~\cite{skinner_measurement-induced_2019,jian_measurement-induced_2020}.
For $\Delta_{\rm cd}=\kappa/2$, spontaneous emission is suppressed, while $\mathcal{E}/\kappa_{\rm eff}\sim\mathcal{O}(1)$. 
In this regime, photon losses continuously extract information about the system through the cavity field. 
The resulting measurement backaction competes with the entanglement generated by the coherent dynamics at a comparable rate, thereby limiting the buildup of long-range entanglement.
For $\Delta_{\rm cd}=5\kappa$, cavity-induced dissipation is suppressed, whereas $\mathcal{E}/\Gamma_{\rm eff}\sim\mathcal{O}(1)$. 
In this case, spontaneous emission provides the dominant monitoring channel and similarly inhibits the growth of long-range entanglement.
These results indicate that observables depending on long-range quantum coherence, such as $S_{N/2}$, are significantly affected by cavity losses and spontaneous emission at $\mathcal{C}$ of a few tens: the two dissipation channels act as monitoring processes limiting accessible entanglement.
The two channels are not however equivalent: for observables linear in the density matrix cavity loss and spontaneous emission act very differently, the former preserving the integrable/chaotic distinction and the latter erasing it. 
Selecting observables of the first class is therefore what keeps chaos signatures accessible at realistic $\mathcal{C}$.

Finally, we estimate the cooperativity $\mathcal{C}$ required to recover the Page value of entanglement entropy for different system sizes, an issue of direct experimental relevance.
Figure~\ref{fig:entanglement_entropy} (d) shows the relative deficit $\delta S_{N/2} = |S_{N/2} - S_{\rm Page}|/S_{\rm Page}$ versus $\mathcal{C}$ at $\Delta_{\rm cd} = 2\kappa$, averaged over trajectories and unitaries~\eqref{eq:f_rqc}. 
At large $\mathcal{C}$ the deficit is compatible with $\sim 1/\sqrt{\mathcal{C}}$ (dashed line), the scaling set by the ratio of the coherent scale to the geometric mean of the dissipative rates, $\mathcal{E}/\sqrt{\kappa_{\rm eff}\Gamma_{\rm eff}} = \sqrt{\mathcal{C}}$. This is the counterpart for analog quantum simulation of the $1/\sqrt{\mathcal{C}}$ infidelity of unconditional two-qubit gates in cavity QED~\cite{sorensen_measurement_2003}. 
It expresses the fact that asymptotically, our structureless many-body dynamics is typically capable maintaining entanglement among $\sim\sqrt{\mathcal{C}}$ atoms.
However, at the realistic $\mathcal{C} \sim 20-200$ the deficit lies well above the $\sqrt{\mathcal{C}}$ line. 
Furthermore, recovering the Page value with a 1$\%$ relative error would demand $\mathcal{C}\sim\mathcal{O}(10^4)$, three orders of magnitude beyond current cavity-QED experiments.

\paragraph{Discussion}
We have investigated chaotic and integrable dynamics in a cavity-QED simulator with ultracold fermions, which provides a promising platform for realizing disordered interacting fermionic models in the laboratory~\cite{uhrich_cavity_2023, baumgartner_quantum_2024, baumgartner_quantum_2025, solis_single-particle_2026, solis2026disorderinducedenhancementfermionicsuperradiance, alves2026densitywaveorderingdisordered}.
Motivated by recent experimental progress in engineering disorder in such systems~\cite{sauerwein_engineering_2023, orsi_cavity_2024}, we considered a protocol designed to generate quantum chaos and examined how dissipation arising from cavity losses and atomic spontaneous emission modifies the corresponding coherent many-body evolution.
We found that the two dissipative mechanisms have qualitatively different structures. Cavity losses reduce to a single effective dephasing channel, whereas, for $\eta\sim\mathcal{O}(1)$, spontaneous emission gives rise to an extensive set of nonlocal dephasing channels.
This distinction identifies a parameter regime in which observables that are linear functionals of the density matrix retain discernible signatures of whether the underlying Hamiltonian dynamics is integrable or chaotic. 
This has a far-reaching implication: despite intrinsic loss channels, a broad class of observables of interest — thermodynamic quantities, transport coefficients, and response functions — can be faithfully accessed by a suitably tuned cavity-based quantum simulator.
Outside this regime, spontaneous emission suppresses these signatures.
By contrast, quantities that depend nonlinearly on the density matrix, such as the entanglement entropy, are substantially more sensitive to decoherence and no longer provide reliable signatures of quantum chaos. 
We interpreted this behavior within the framework of monitored quantum many-body dynamics.

More generally, our results show that dissipation can qualitatively modify the behavior expected from the corresponding unitary dynamics in cavity-QED systems designed to generate spatially extended entangled states. 
Achieving faithful control of many-body chaotic dynamics requires very large cooperativities.
Nevertheless, at experimentally realistic values of $\mathcal{C}$, selected signatures of quantum chaos may remain observable, provided that the measured quantities and operating parameters are chosen appropriately. 
Although the present work focuses on quantum chaos, the effects of atomic spontaneous emission are also expected to be relevant to other resonator-mediated many-body phenomena.

\begin{acknowledgments}
\paragraph{Acknowledgements}
We acknowledge useful discussions with A. Paviglianiti, A. Mercurio, M. Seclì, F. Minganti, L. Goutte, L. Fioroni, E. Zhao and E. Tirrito.
This work was supported by the Swiss National Science Foundation through Projects No. 200020\_215172 and 217124, by the Swiss State Secretariat for Education, Research and Innovation (Grants No. MB22.00063 and 20QU-1\_215924) and by NCCR Precision, a National Centre of Competence in Research funded by the Swiss National Science Foundation grant 51AU-0\_229299. 
The authors would like to thank the Swiss National Science Foundation for their financial support. 
All numerical simulations are obtained with the \texttt{QuantumToolbox.jl} package \cite{mercurio_quantum_2025}.
\end{acknowledgments}


%

\vspace{2cm}

\onecolumngrid
\begin{center}
  \textbf{\large End Matter}\\[1em]
\end{center}
\vspace{1em}
\twocolumngrid
\paragraph{Quantum many-body dynamics in the Lamb-Dicke regime}
\begin{figure}[t]
\centering
\includegraphics[width=0.5\textwidth]{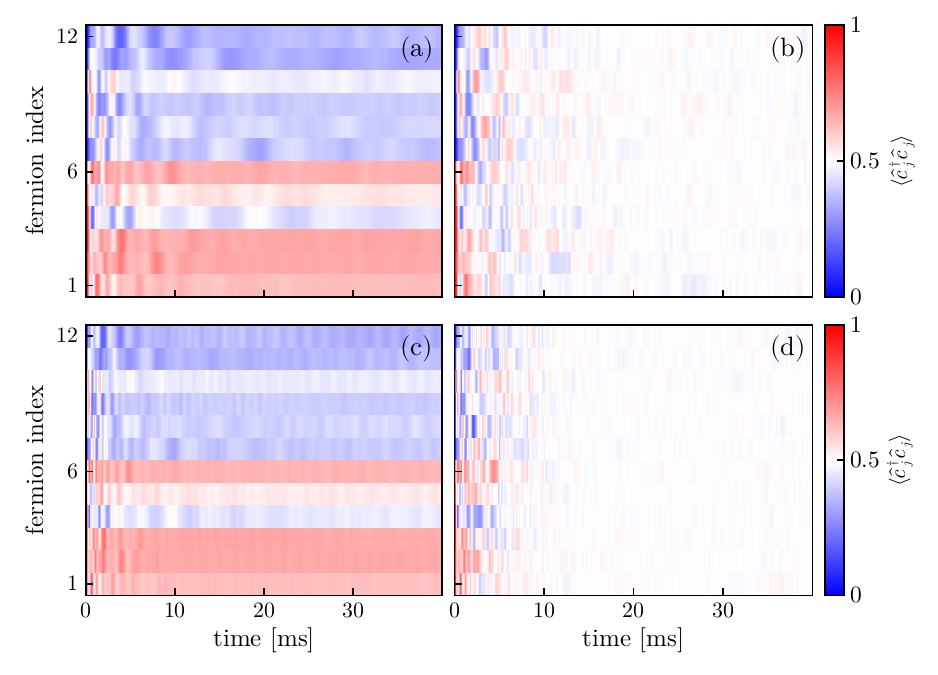}\vspace{0.05em}
\caption{
Quantum many-body dynamics with $\eta\ll1$ for $\Delta_{\rmc\rmd}=\kappa/2$ (first row) and $\Delta_{\rmc\rmd}=5\kappa$ (second row).
The first column refers to the static disorder configuration. 
The second column to the f-RQC protocol.
The harmonic oscillator length is fixed to $x_0=5\,\textrm{nm}$ such that $\eta=0.047$.
All the parameters are fixed as in Fig.~\ref{fig:small_detuning}.
}
\label{fig:small_lamb_dicke}
\end{figure}
To demonstrate that the detrimental action of spontaneous emission on the quantum many-body dynamics is precisely due to the proliferation of dephasing jump operators due to recoil effects at sizable $\eta$, we consider here a configuration deep in the Lamb-Dicke regime.
Since $\lambda_\rmc$ is fixed, we decrease the oscillator length to $x_0=5\,\textrm{nm}$, obtaining $\eta=0.047$.
We remark that such a configuration is not experimentally feasible for an optical wavelength of the hundreds of micrometers due to diffraction limits~\cite{born_principles_1999}, and it serves to illustrate the qualitative difference between $\eta\sim\mathcal{O}(1)$ and $\eta\ll1$. 
We note that the coherent, photon-mediated couplings responsible for the chaotic dynamics originate from the spatial overlap of the orbitals and the speckle pattern and do not vanish as $\eta \to 0$; only the recoil-induced heating channel is suppressed in that limit~\suppinfo.

In Fig.~\ref{fig:small_lamb_dicke} we plot the dynamics of $\langle\hat{c}^\dagger_j\hat{c}_j\rangle$ for $N=12$ fermionic modes at half filling.
The first row displays the data for $\Delta_{\rm cd}=\kappa/2$, while the second row the data for $\Delta_{\rm cd}=5\kappa$.
Columns refer to the static and dynamical disorder configurations, respectively.
We observe how, in the Lamb-Dicke regime, the distinction between integrable and chaotic dynamics is possible throughout all the values of $\Delta_{\rmc\rmd}$ considered.
In other words, the destructive effect of spontaneous emission on the coherent dynamics of the system is suppressed by $\eta$.
This qualitative difference with respect to the experimentally relevant case $\eta\sim\mathcal{O}(1)$ discussed in the main text is due to how the structure of the spontaneous emission dissipator changes with $\eta$.
Away from the Lamb-Dicke regime, the Lindblad dissipator associated with spontaneous emission is given by Eq.~\eqref{eq:spontaneous_emission_effective_dissipator}, where $K_{jk\ell m}$ has a high-rank structure.
This leads to a proliferation of dephasing jump operators encoding the physics of recoil and heating.
If $\eta\ll1$, instead, we obtain (see the Supplementary Information)
\begin{equation}\label{eq:spontaneous_emission_effective_dissipator_lamb_dicke}
    \mathcal{D}\hat{\rho} \sim \Gamma_{\rm eff} \sum_{jk\ell m}K_{jk}K_{\ell m}\hat{c}_j^\dagger\hat{c}_k\,\hat{\rho}\,\hat{c}_\ell^\dagger\hat{c}_m.
\end{equation}
The structure of the spontaneous emission dissipator now coincides with the one associated with cavity dissipation in Eq.~\eqref{eq:cavity_effective_dissipator}.
Thus, Eq.~\eqref{eq:spontaneous_dissipator_lambd_dicke} yields a \textit{single} Lindblad jump operator $\hat{L}_\rma = \sqrt{\Gamma_{\rm eff}}\sum_{jk}K_{jk}\hat{c}_j^\dagger\hat{c}_k$ with $|K_{jj}|\gg|K_{j\ne k}|$.
This peculiar structure emerging in the Lamb-Dicke regime keeps the effects of dissipation negligible on the averaged quantum many-body dynamics.
The Lamb-Dicke regime is therefore beneficial in principle, even though it is not accessible at optical wavelengths.

\paragraph{Effects of the trap Hamiltonian}
Our treatment in the main text neglected the trap Hamiltonian which sums to Eq.~\eqref{eq:hamiltonian} \suppinfo.
In this way, our discussion was completely general and anchored on only three ingredients: the cavity single-atom cooperativity $\mathcal{C}$, the fixed cavity dissipation rate $\kappa/2\pi=200\,\textrm{kHz}$, the dispersive regime between the pump and the cavity.
Our results are independent of the atomic species in a cold atom experiment and they essentially rely on the strength of the light-matter interaction versus decoherence processes.
Indeed, even though we assumed to work with $^6$Li atoms, having chosen $\Gamma/2\pi=5.9\,\textrm{MHz}$, we considered $\mathcal{C}=20$, which fixed $\Omega$ and we assumed to be in the dispersive regime ($\Omega\Omega_\rmd/\Delta_{\rmc\rmd}\Delta_{\rmd\rma}\simeq 0.2$), which fixed all the other tunable parameters. 
As such, changing species does not modify the various effective time scales in Eqs.~\eqref{eqs:time_scales}, nor affects the physics we explored.

We now want to study the effects of the trap Hamiltonian, describing a non-interacting (thus integrable) contribution coming from simply having fermions in a 2D harmonic trap, and it amounts into the term
\begin{equation}\label{eq:harmonic_trap}
    \hat{H}_{\rm trap} = \sum_j\varepsilon_j\hat{c}_j^\dagger\hat{c}_j,
\end{equation}
where $\varepsilon_j = \omega_\rmt(n_j+1)$, being $n_j = n_j^{(x)}+n_j^{(y)}$, and $\omega_\rmt=\hbar/m_{\rm at}x_0^2$, where $m_{\rm at}$ is the atomic mass.
The trapping frequency $\omega_\rmt$ fixes the typical energy scale of Eq.~\eqref{eq:harmonic_trap}.
In this paragraph we work at $x_0=300\,\textrm{nm}$, such that the trapping frequency is fully determined by $m_{\rm at}$.
We consider two fermionic isotopes, that are largely used by the cold atom community: $^6$Li (dipole transition) and $^{171}$Yb (intercombination transition).
Relevant parameters are reported in Table~\ref{table:yb_li}.

\begin{table}[h!]
\centering
\begin{tabular}{lcccc}
\toprule
Atom\quad &  \quad$\lambda_\rmc$\quad & \quad$\Gamma/2\pi$\quad & \quad$\omega_\rmt/2\pi$\quad & \quad$\eta$\quad \\
\midrule
$^6$Li\qquad  & \quad671 nm\quad & \quad5.9 MHz\quad   & \quad$19\,\textrm{kHz}$\quad  & \quad$2.81$\quad \\
$^{171}$Yb\quad & \quad556 nm\quad & \quad182 kHz\quad  & \quad$0.6 \,\textrm{kHz}$\quad & \quad$3.39$\quad \\
\bottomrule
\end{tabular}
\caption{Transition wavelength and linewidth, trapping frequency at fixed $x_0=300\,\textrm{nm}$ and Lamb-Dicke parameter for $^6$Li atoms (first row) and $^{171}$Yb atoms (second row).}
\label{table:yb_li}
\end{table}

\begin{figure}[t]
\centering
\includegraphics[width=0.5\textwidth]{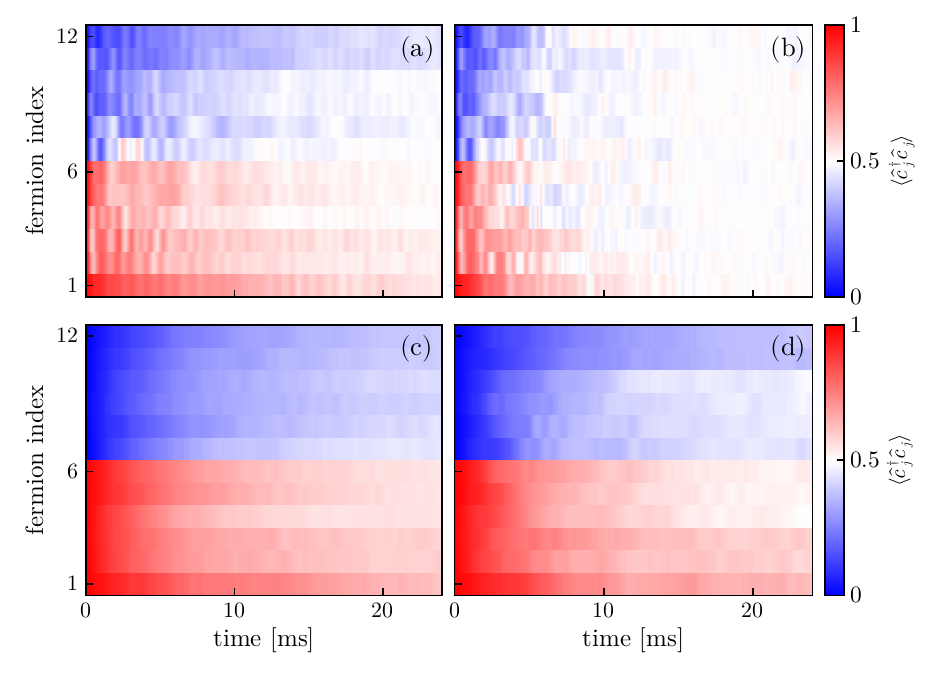}\vspace{0.05em}
\caption{
Effects of the trap Hamiltonian for $^{171}$Yb atoms (first row) and $^6$Li atoms (second row) on the quantum many-body dynamics.
The first column refers to the static disorder configuration. 
The second column to the f-RQC protocol.
All the parameters are fixed as in Fig.~\ref{fig:small_detuning}.
}
\label{fig:trap_hamiltonian}
\end{figure}

The presence of the trap Hamiltonian now forces the comparison between the effective long-range interaction strength $\mathcal{E}$, yielding the desired physics, and $\omega_\rmt$.
For the parameters reported in Fig.~\ref{eq:hamiltonian_effective} the long-range interaction strength is equal to $\mathcal{E}/2\pi\simeq4\,\textrm{kHz}$. 
We immediately realize that $\mathcal{E}\ll\omega_\rmt$ for harmonically trapped $^6$Li in an optical tweezer, while $\mathcal{E}\gtrsim\omega_\rmt$ for $^{171}$Yb atoms.

To illustrate the unwanted effects associated with the trap Hamiltonian we study the quantum many-body dynamics of $\langle\hat{c}_j^\dagger\hat{c}_j\rangle$ generated by Eq.~\eqref{eq:hamiltonian} including the trap Hamiltonian, cavity dissipation and atomic spontaneous emission, for $^{171}$Yb and $^{6}$Li atoms.
Results are reported in Fig.~\ref{fig:trap_hamiltonian}.
Panels (a) and (b) display the dynamics of $\langle\hat{c}_j^\dagger\hat{c}_j\rangle$ for $^{171}$Yb atoms subject to static disorder and the f-RQC protocol, respectively.
In this case, the trapping frequency is smaller than the long-range interaction strength and the dynamics is very similar to the one reported in Fig.~\ref{fig:small_detuning} (a) and (b), where no trap Hamiltonian is considered.
Panels (c) and (d) shows $\langle\hat{c}_j^\dagger\hat{c}_j\rangle$ as a function of time for $^6$Li atoms with the same disorder schemes used in panels (a) and (b).
Here, we have a much larger $\omega_\rmt$, as reported in Table \ref{table:yb_li}.
The integrable trap term in Eq.~\eqref{eq:hamiltonian} therefore dominates the dynamics in the milliseconds range effectively freezing the fermionic densities in the initial configuration.
Regardless of the disorder protocol, either static or dynamical, the system displays evident integrability fingerprints, and the entanglement entropy stays close to zero (not shown).
This analysis highlights the detrimental effect of a large trapping frequency $\omega_\rmt$, which grows both for a tightly confining trap (small $x_0$) and for a light atomic species (small $m_{\rm at}$).
When $\omega_\rmt\gg\mathcal{E}$, the integrable trap term dominates and freezes the fermionic densities in their initial configuration.

A careful engineering of the trapping potential could be advantageous for the control of many-body quantum chaos in cavity-QED architectures, and we argue that it would simultaneously mitigate the effect of the trap Hamiltonian and the heating induced by spontaneous emission.
The two goals are best served not by a uniformly shallower trap--which would only increase $x_0$, and hence the Lamb–Dicke parameter $\eta=k_0x_0$, worsening the recoil--but by a tailored level structure: a dense manifold of closely spaced levels, with spacing below $\mathcal{E}$ so that the interaction can drive the dynamics, separated by a large gap from the higher, unpopulated levels. 
Such a spectrum decouples the intra-manifold spacing from the size of the gap, a freedom that a harmonic trap does not offer. 
An optical lattice realizes both features at once: the lowest band furnishes the dense working manifold while the band gap protects it from above.
Notably, such an energy gap would also control the validity of the finite-orbital approximation assumed in this paper.
It would prevent heating and momentum recoil from ejecting the atoms outside the fixed orbital manifold, a condition that we assumed to remain valid in all our numerical simulations.

\clearpage

\clearpage
\onecolumngrid

\begin{center}
  \textbf{\large Supplementary Information for \\ ``Controlling many-body quantum chaos in a dissipative optical cavity''}\\[1em]
  Filippo Ferrari, Francesca Orsi, Ekaterina Fedotova, Oscar Rios Alves, \\Michał Zdziennicki, Jean-Philippe Brantut, Vincenzo Savona
\end{center}
\vspace{1em}

\titleformat{\section}{\bfseries\centering}{\thesection.}{1em}{}
\titleformat{\subsection}{\bfseries\centering}{\thesubsection}{1em}{}
\titleformat{\subsubsection}{\centering\itshape}{\thesubsubsection}{1em}{}

\renewcommand{\thesection}{\Roman{section}}
\renewcommand{\thesubsection}{\Alph{subsection}}
\renewcommand{\thesubsubsection}{\Alph{subsection}}

\section{I. DERIVATION OF THE DRIVEN-DISSIPATIVE FERMION-CAVITY SYSTEM}

We derive here the Hamiltonian~\eqref{eq:hamiltonian} in the main text, together with the collection of spontaneous emission jump operators used in the numerical simulations.

\subsection{A. Hamiltonian}

\subsubsection{1. Adiabatic elimination of the atomic excited states}

We consider an ensemble of fermionic atoms harmonically trapped within a single-mode optical cavity. 
The full many-body Hamiltonian in the rotating wave approximation reads
\begin{equation}\label{eq:many_body}
\begin{split}
    \hat{H} &= \omega_\rmc\hat{a}^\dagger\hat{a} + \sum_{\rms=\rme,\,\rmg}\int\rmd^2r\,\hat{\Psi}_\rms^\dagger(\bfr)\left[-\frac{\nabla^2}{2m_{\rm at}} + V(\bfr)\right]\hat{\Psi}_\rms(\bfr) + \int\rmd^2 r\,\omega_\rma(\bfr)\hat{\Psi}_\rme^\dagger(\bfr)\hat{\Psi}_\rme(\bfr)\\& + \frac{\Omega}{2}\int\rmd^2 r\,g(\bfr)\left[\hat{a}\,\hat{\Psi}_\rme^\dagger(\bfr)\hat{\Psi}_\rmg(\bfr) + \hat{a}^\dagger\,\hat{\Psi}_\rmg^\dagger(\bfr)\hat{\Psi}_\rme(\bfr)\right] + \Omega_\rmd \int \rmd^2 r\,g_\rmd(\bfr)\left[\rme^{-\rmi\omega_\rmd t}\,\hat{\Psi}_\rme^\dagger(\bfr)\hat{\Psi}_\rmg(\bfr) + \rme^{+\rmi\omega_\rmd t}\,\hat{\Psi}_\rmg^\dagger(\bfr)\hat{\Psi}_\rme(\bfr)\right].
\end{split}
\end{equation}
Where we have, in order: i) the cavity at frequency $\omega_\rmc$, ii) the fermions in the harmonic trap with potential $V(\bfr) = \frac{1}{2}m_{\rm at}\omega_\rmt^2\bfr^2$ with trap frequency $\omega_\rmt=\hbar/m_{\rm at}x_0^2$, being $m_{\rm at}$ the atomic mass and $x_0$ the harmonic oscillator length, iii) the excited state energy at (space dependent) frequency $\omega_\rma(\bfr)$, iv) the light-matter couplings with coupling strength (or bare Rabi amplitude) $\Omega$, and v) the drive on the atoms with amplitude $\Omega_\rmd$.
In Eq.~\eqref{eq:many_body}, we describe the fermions considering the space-dependent fields operators $\hat{\Psi}_\rme^\dagger(\bfr)$ and $\hat{\Psi}_\rmg^\dagger(\bfr)$ for the excited and ground atomic states, respectively.
Finally, $g(\bfr)$ and $g_\rmd(\bfr)$ are the classical drive and cavity lattice, that we assume without loss of generality to be real-valued.
We first operate a unitary transformation by moving to the rotating frame at $\omega_\rmd$, $\hat{U} = \textrm{exp}(-\rmi\hat{H}_{\rm RF}t)$ with
\begin{equation}
   \hat{H}_{\rm RF} = \omega_\rmd\int \rmd^2 r\hat{\Psi}_\rme^\dagger(\bfr)\hat{\Psi}_\rme(\bfr) + \omega_\rmd\hat{a}^\dagger\hat{a},
\end{equation}
to get the time-independent Hamiltonian
\begin{equation}
\begin{split}
     \hat{H} &= \Delta_{\rmc\rmd}\hat{a}^\dagger\hat{a} + \sum_{\rms=\rme,\,\rmg}\int\rmd^2r\,\hat{\Psi}_\rms^\dagger(\bfr)\left[-\frac{\nabla^2}{2m_{\rm at}} + V(\bfr)\right]\hat{\Psi}_\rms(\bfr) - \int\rmd^2 r\Delta_{\rmd\rma}(\bfr)\hat{\Psi}_\rme^\dagger(\bfr)\hat{\Psi}_\rme(\bfr) \\&+ \int\rmd^2 r\left[\hat{\Phi}(\bfr)\hat{\Psi}_\rme^\dagger(\bfr)\hat{\Psi}_\rmg(\bfr) + \hat{\Phi}^\dagger(\bfr)\hat{\Psi}_\rmg^\dagger(\bfr)\hat{\Psi}_\rme(\bfr)\right],
\end{split}
\end{equation}
where $\Delta_{\rmc\rmd}=\omega_c-\omega_d$ is the cavity-to-drive detuning, $\Delta_{\rmd\rma}(\bfr)=\omega_\rmd-\omega_\rma(\bfr)$ is the drive-to-atom detuning, and $\hat{\Phi}(\bfr) = \Omega_\rmd g_\rmd(\bfr) + \frac{\Omega}{2}g(\bfr)\hat{a}$.
We now perform the elimination of the atomic excited states, motivated by the fact that $|\Delta_{\rmd\rma}(\bfr)|$ is the dominant energy scale and the atoms are deep in the dispersive regime with respect to the drive and cavity resonances, meaning that $\hat{\Psi}_\rme(\bfr)$ adiabatically follows $\hat{\Psi}_\rmg(\bfr)$.
The adiabatic condition can be obtained from the Heisenberg equation for $\hat{\Psi}_\rme(\bfr)$,
\begin{equation}
    0 = \frac{\partial \hat{\Psi}_\rme(\bfr)}{\partial t} = \rmi[\hat{H}, \hat{\Psi}_\rme(\bfr)] = -\rmi\Delta_{\rmd\rma}\hat{\Psi}_\rme(\bfr) + \rmi\hat{\Phi}(\bfr)\hat{\Psi}_\rmg(\bfr),
\end{equation}
which yields $\hat{\Psi}_\rme(\bfr) = \hat{\Phi}(\bfr)\hat{\Psi}_\rmg(\bfr)/\Delta_{\rmd\rma}$.
The final Hamiltonian, after the adiabatic elimination, reads
\begin{equation}
\begin{split}
    \hat{H} =
    \Delta_{\rmc\rmd}\hat{a}^\dagger\hat{a} + \int\rmd^2r\,\hat{\Psi}_\rmg^\dagger(\bfr)\left[-\frac{\nabla^2}{2m_{\rm at}} + V(\bfr)\right]\hat{\Psi}_\rmg(\bfr) + \int\frac{\rmd^2 r}{\Delta_{\rmd\rma}(\bfr)}\hat{\Phi}^\dagger(\bfr)\hat{\Phi}(\bfr) \hat{\Psi}_\rmg^\dagger(\bfr)\hat{\Psi}_\rmg(\bfr).
\end{split}
\end{equation}
Notice that in the above equation we dropped the term proportional to $\omega_\rmt$ and containing the atomic excited states, as it depends on $\Delta^2_{\rmd\rma}(\bfr)$ at the denominator.

\subsubsection{2. Single-particle wave functions}

We now expand the fermionic field operator $\hat{\Psi}_\rmg(\bfr)$ in single-particle wave functions and many-body annihilation operators, $\hat{\Psi}_\rmg(\bfr) = \sum_{j=1}^N\varphi_j(\bfr)\,\hat{c}_j$.
The single-particle wave functions depends on the trapping geometry.
In this case, we assume a 2D harmonic trap such that $\varphi_j(\bfr) = \psi_{n_x}(x)\psi_{n_y}(y)$ and
\begin{equation}
    \psi_n(x) = \sqrt{\frac{1}{\sqrt{\pi}\,2^n\,n!\,x_0}}\,\rme^{-x^2/2x_0^2}\,H_n\left(\frac{x}{x_0}\right),
\end{equation}
where $H_n(z)$ are the Hermite polynomials.
In the numerical simulations, we work on a 2D space grid $[-L/2, L/2]\times[-L/2, L/2]$ with $L/x_0=15$ and we use $N_L=500$ points per side to resolve the single-particle wavefunctions.
The many-body annihilation operators encodes the fermionic anti-commutation rules,
\begin{equation}
    \{\hat{c}_j, \hat{c}_k^\dagger\} = \delta_{jk}\mathds{1},\qquad \{\hat{c}_j, \hat{c}_k\} = \{\hat{c}_j^\dagger, \hat{c}_k^\dagger\} = 0.
\end{equation}
In the numerical simulations, $\hat{c}_j$ and $\hat{c}_j^\dagger$ are constructed by means of a Jordan-Wigner transformation and they are $2^N\times 2^N$ matrices. 
Since all the fermionic Hamiltonians we study across this article commute with the number operator $\hat{N} = \sum_{j=1}^N\hat{c}_j^\dagger\hat{c}_j$, we can restrict ourselves to the half-filling subspace with $N/2$ particles and $N$ states.
The size of the matrices describing fermionic operators thus reduces to $\binom{N}{N/2}\times\binom{N}{N/2}$.

\subsubsection{3. Generation of the speckle potential}

The disorder in the system is implemented via the randomization of the drive-atom detuning $\Delta_{\rmd\rma}(\bfr)$.
In practice, the cloud of fermionic atoms is subject to a spatially disordered AC-Stark shift that off-resonantly dresses the excited state $\ket{\rme}$ with an auxiliary state $\ket{\rma}$.
This shifts the excited state by an amount $|\Omega_{\rmb}(\bfr)|^2/4\Delta_{\rmb}$. This implies
\begin{equation}
    \Delta_{\rmd\rma}(\bfr) = \omega_\rmd - \omega_\rma(\bfr) = (\omega_\rmd - \omega_\rma) - \frac{|\Omega_\rmb(\bfr)|^2}{4\Delta_\rmb} = \Delta_{\rmd\rma} - \frac{|\Omega_\rmb(\bfr)|^2}{4\Delta_\rmb}.
\end{equation}
The disorder is implemented through an optical speckle pattern $I(\bfr)\propto|\Omega_\rmb(\bfr)|^2$.
To generate the speckle potential we proceed as follows.
We start with the space grid with side $L$ containing $N_L\times N_L$ points.
We then define the frequency grid as

\begin{equation}
    k_{x,j} = \begin{cases}
\frac{2\pi j}{L} & \text{for } j = 0, \ldots, N_L/2 - 1, \\
\frac{2\pi (j - N_L)}{L} & \text{for } i = N_L/2, \ldots, N_L - 1,
\end{cases}
\end{equation}
and same for $k_{y,j}$. 
This defines the wave vector magnitude $K(k_x, k_y) = \sqrt{k^2_x+ k^2_y}$.
We then define the random phase field in $k$-space as $\phi(k_x, k_y) = 2\pi\times\textrm{rand}(0, 1)$.
Next, we choose a sharp cutoff $k_{\rm max} = 2\pi/\xi$, being $\xi$ the correlation length of the speckle pattern.
This implies
\begin{equation}
    F(K) = \begin{cases}
1 & \text{if } K < k_{\text{max}}, \\
0 & \text{otherwise}.
\end{cases}
\end{equation}
We then construct the complex field in $k$-space as $\tilde{E}(k_x, k_y) = F(K)\,\rme^{\rmi\phi(k_x, k_y)}$.
This is a superposition of plane waves with random phases and uniform amplitude if $K<k_{\rm max}$.
To pass in real space we Fourier-invert the complex field to get $E(x, y) = \mathcal{F}^{-1}[\tilde{E}(k_x, k_y)]$, we compute the intensity $I(\bfr) = |E(\bfr)|^2$ and we normalize it.
Therefore, the detuning becomes
\begin{equation}
    \Delta_{\rmd\rma}(\bfr) = \Delta_{\rmd\rma}\left[1+f\frac{I(\bfr)}{\langle I \rangle}\right],
\end{equation}
where we introduced the parameter $f = \langle I \rangle / (4\Delta_\rmb\Delta_{\rmd\rma})$ that controls the disorder strength.
In the numerical simulations presented in the paper, we assumed $f=1$, that assumes a strong disorder configuration, and $\xi=4x_0$ for the speckle correlation length.

\subsubsection{4. Hamiltonian in second-quantization}

Finally, we explicitly express the Hamiltonian in second quantization, obtaining Eq.~\eqref{eq:hamiltonian} in the main text. 
We get indeed
\begin{equation}
\begin{split}
    \hat{H} &= \Delta_{\rmc\rmd}\hat{a}^\dagger\hat{a} +  \int\rmd^2r\,\hat{\Psi}_\rmg^\dagger(\bfr)\left[-\frac{\nabla^2}{2m_{\rm at}} + V(\bfr)\right]\hat{\Psi}_\rmg(\bfr) + \Omega_\rmd^2\int\rmd^2 r\frac{g_\rmd^2(\bfr)}{\Delta_{\rmd\rma}(\bfr)}\hat{\Psi}_\rmg^\dagger(\bfr)\hat{\Psi}_\rmg(\bfr) \\&+ \frac{\Omega^2}{4}\int\rmd^2 r\frac{g^2(\bfr)}{\Delta_{\rmd\rma}(\bfr)}\,\hat{a}^\dagger\hat{a}\,\hat{\Psi}_\rmg^\dagger(\bfr)\hat{\Psi}_\rmg(\bfr) + \frac{\Omega\Omega_\rmd}{2}\int\rmd^2 r\frac{g_\rmd(\bfr)g(\bfr)}{\Delta_{\rmd\rma}(\bfr)}\,(\hat{a}^\dagger+\hat{a})\,\hat{\Psi}^\dagger_\rmg(\bfr)\hat{\Psi}_\rmg(\bfr).
\end{split}
\end{equation}
The first term is the cavity. 
The second term is the trap Hamiltonian for the atomic ground states.
The third term is a dipole trap generated by the pump, that can be compensated with an additional drive, as explained in Ref.~\cite{uhrich_cavity_2023}.
As such, we will ignore it throughout the paper.
The fourth term in the dispersive regime amounts for a detuning renormalization.
The last term gives the desired interaction coefficients responsible for the all-to-all random couplings in Eq.~\eqref{eq:hamiltonian_effective}.
We now drop the term depending on the fermions only and we expand the field operators to get Eq.~\eqref{eq:hamiltonian}
\begin{equation}\label{eq:second_quantized}
    \hat{H} = \Delta_{\rmc\rmd}\hat{a}^\dagger\hat{a} + \sum_j\varepsilon_j\,\hat{c}_j^\dagger\hat{c}_j + \frac{\Omega^2}{4\Delta_{\rmd\rma}}\sum_{jk}d_{jk}\,\hat{a}^\dagger\hat{a}\,\hat{c}_j^\dagger\hat{c}_k + \frac{\Omega\Omega_\rmd}{2\Delta_{\rmd\rma}}\sum_{jk}g_{jk}\,(\hat{a}^\dagger + \hat{a})\,\hat{c}_j^\dagger\hat{c}_k,
\end{equation}
where the diagonal trapping energies are
\begin{equation}
    \varepsilon_j = \omega_\rmt\left(n_j+1\right),
\end{equation}
where $n_j = n_x^{(j)} + n_y^{(j)}$ are ordered from below,
and the all-to-all random couplings are
\begin{equation}\label{eqs:random_coefficients}
    d_{jk} = \int\rmd^2 r\,\frac{g^2(\bfr)\varphi_j(\bfr)\varphi_k(\bfr)}{1+I(\bfr)/\langle I \rangle},\qquad g_{jk} = \int\rmd^2 r\,\frac{g_\rmd(\bfr)g(\bfr)\varphi_j(\bfr)\varphi_k(\bfr)}{1+I(\bfr)/\langle I \rangle}.
\end{equation}
In the numerical simulations, we will also neglect the photon number-dependent quadratic fermionic Hamiltonian.
This is justified since $\Omega_\rmd\gg\Omega$ and $\langle\hat{a}^\dagger\hat{a}\rangle\simeq0$ (dispersive regime between the cavity and the drive).
Finally, we simplify the random coefficients $g_{jk}$ in Eqs.~\eqref{eqs:random_coefficients} by assuming $g(\bfr)=g_\rmd(\bfr)\simeq1$.
This coincides with having cavity and drive beam waists much larger than the oscillator length $x_0$.
We can not exclude \textit{a priori} that a precise engineering of the classical cavity and drive fields could improve the results presented in the main text.
However, we leave this optimization task for future work.

\subsection{B. Atomic spontaneous emission}

Atomic spontaneous emission is the process by means an atom in the excited state decay in the ground state emitting a photon at wavevector $k_0=2\pi/\lambda_\rmc$, being $\lambda_\rmc$ the cavity wavelength.
The microscopic description of this dissipative effect in a many-body setting starts by coupling $\hat{\Psi}_\rmg^\dagger\hat{\Psi}_\rme$ to a continuum of bosonic vacuum modes at wavevectors $\bfk$ and tracing them out~\cite{daley_quantum_2014}. 
This yields a spontaneous emission Lindblad dissipator built from one jump operator per emission direction $\bfn$~\cite{pichler_nonequilibrium_2010}.
Because the atoms are frozen in the plane (tight
$z$ confinement), only the in-plane projection $\bfn_\perp$ of the photon momentum couples to the
orbital motion, and each jump operator carries the recoil $k_0\bfn_\perp$:
\begin{equation}\label{eq:n_jump_operator}
    \hat{L}_\rma(\bfn) = \sqrt{\Gamma}\int \rmd^2r\;
  \rme^{-\rmi k_0\,\bfn_\perp\cdot\bfr}\,\hat{\Psi}_\rmg^\dagger(\bfr)\hat{\Psi}_\rme(\bfr).
\end{equation}
Even if the atoms are confined within a 2D geometry, they radiate into the full solid angle.
Writing $\bfn=(\sin\Theta\cos\phi,\,\sin\Theta\sin\phi,\,\cos\Theta)$, the
in-plane projection is $\bfn_\perp=\sin\Theta\,(\cos\phi,\,\sin\phi)$, while the out-of-plane recoil $k_0\cos\Theta$ acts on the tightly confined $z$ mode and is frozen out. 
For simplicity, we consider isotropic (scalar) emission.
The spontaneous emission Lindblad dissipator results from the sum over all the possible emission directions and reads~\cite{pichler_nonequilibrium_2010}
\begin{equation}
    \mathcal{D}[\hat{L}_\rma(\bfn)]\hat{\rho} = \int\frac{\rmd\Omega_\bfn}{4\pi}\,\left[\hat{L}_\rma(\bfn)\,\hat{\rho}\,\hat{L}^\dagger_\rma(\bfn)-\frac{1}{2}\{\hat{L}^\dagger_\rma(\bfn)\hat{L}_\rma(\bfn),\,\hat{\rho}\}\right],
\end{equation}
which is equivalent to
\begin{equation}\label{eq:spontaneous_emission_general}
\begin{split}
    \mathcal{D}[\hat{L}_\rma]\hat{\rho} &= \Gamma\int\rmd^2r\rmd^2r'\,j_0(k_0|\bfr-\bfr'|)\left[\hat{\Psi}_\rmg^\dagger(\bfr)\hat{\Psi}_\rme(\bfr)\,\hat{\rho}\,\hat{\Psi}_\rme^\dagger(\bfr')\hat{\Psi}_\rmg(\bfr')-\frac{1}{2}\{\hat{\Psi}_\rme^\dagger(\bfr')\hat{\Psi}_\rmg(\bfr')\hat{\Psi}_\rmg^\dagger(\bfr)\hat{\Psi}_\rme(\bfr),\,\hat{\rho}\}\right]\\&
    =\int\rmd^2r\rmd^2r'\,j_0(k_0|\bfr-\bfr'|)\left[\hat{L}_\rma(\bfr)\,\hat{\rho}\,\hat{L}^\dagger_\rma(\bfr')-\frac{1}{2}\{\hat{L}^\dagger_\rma(\bfr')\hat{L}_\rma(\bfr),\,\hat{\rho}\}\right],
\end{split}
\end{equation}
since $\int\frac{\rmd\Omega_\bfn}{4\pi}\,\mathrm{e}^{-ik_0\,\bfn_\perp\cdot(\bfr-\bfr')}=\int_0^{\pi}\frac{\sin\Theta\,\rmd\Theta}{2}\,\mathcal{J}_0\!\bigl(k_0 \sin\Theta |\bfr-\bfr'|\bigr) = j_0(k_0|\bfr-\bfr'|)$.
We also re-define $\hat{L}_\rma = \sqrt{\Gamma}\hat{\Psi}_\rmg^\dagger(\bfr)\hat{\Psi}_\rme(\bfr)$. 
Here, $\mathcal J_0(x)$ is the Bessel function of the first kind and $j_0(x)=\sin(x)/x$ is the spherical Bessel function of the first kind. 
It satisfies $j_0(0)=1$, which fixes the total scattering
rate, and varies on the scale $k_0^{-1}=\lambda_\rmc/2\pi$. 
The spherical Bessel function kernel is called photon-coherence kernel for 3D scalar emission.
It measures the distance over which emission from different points stays mutually coherent.
The dimensionless ratio of this coherence length to the orbital size is the so-called Lamb-Dicke parameter $\eta=k_0x_0 = 2\pi x_0/\lambda_\rmc$.
While Eq.~\eqref{eq:spontaneous_emission_general} assumes scalar radiation, a more realistic description would imply a dipole pattern, that can be captured by replacing the uniform average $\int\frac{\rmd\Omega_\bfn}{4\pi}$ with the weighted average $\int\frac{\rmd\Omega_\bfn}{4\pi}(1-|\bfd\cdot\bfn|^2)$, being $\bfd$ the atomic dipole momentum.
This however does not change the $\sim\lambda_\rmc$ coherence length of the emitted photons.

As Eq.~\ref{eq:spontaneous_emission_general} involves both excited and ground atomic states, since $|\Delta_{\rmd\rma}(\bfr)|\gg\Gamma,\Omega_\rmd$ we can apply the adiabatic elimination of the excited states.
From the Heisenberg equations of motion it follows that
\begin{equation}
    \hat{\Psi}_\rmg^\dagger(\bfr)\hat{\Psi}_\rme(\bfr) = \frac{\hat{\Phi}(\bfr) \hat{\Psi}_\rmg^\dagger(\bfr)\hat{\Psi}_\rmg(\bfr)}{\Delta_{\rmd\rma}(\bfr) + \rmi\Gamma/2}\simeq\frac{\hat{\Phi}(\bfr) \hat{\Psi}_\rmg^\dagger(\bfr)\hat{\Psi}_\rmg(\bfr)}{\Delta_{\rmd\rma}(\bfr)}.
\end{equation}
And we deduce the final form of the space-dependent jump operator
\begin{equation}
    \hat{L}_\rma(\bfr) = \frac{\sqrt{\Gamma}}{\Delta_{\rmd\rma}(\bfr)}\left[\Omega_\rmd g_\rmd(\bfr) + \frac{\Omega}{2}g(\bfr)\hat{a}\right]\hat{\Psi}_\rmg^\dagger(\bfr)\hat{\Psi}_\rmg(\bfr)\simeq\sqrt{\Gamma}\frac{\Omega_\rmd g_\rmd(\bfr)}{\Delta_{\rmd\rma}(\bfr)}\hat{\Psi}_\rmg^\dagger(\bfr)\hat{\Psi}_\rmg(\bfr),
\end{equation}
where in the last passage we neglect the contribution coming from the cavity field, an approximation valid in the dispersive regime where the cavity is close to the vacuum and if $\Omega_\rmd\gg\Omega$.
Equivalently, we are considering the simplest 0-th order contribution from $\mathcal{D}[\hat{L}_\rma]\hat{\rho}$ in powers of $\hat{a}$ and $\hat{a}^\dagger$.
This amounts in neglecting cavity-assisted spontaneous emission.
We can now expand the dissipator in single-particle wavefunctions and fermionic creation and annihilation operators.
We get
\begin{equation}\label{eq:spontaneous_dissipator}
\begin{split}
    \int\rmd^2r\rmd^2r'\, j_0(k_0|\bfr-\bfr'|)\,\hat{L}_\rma(\bfr)\hat{\rho}\hat{L}^\dagger_\rma(\bfr') 
    &=
    \Gamma\Omega_\rmd^2 \int\rmd^2r\rmd^2r'\, j_0(k_0|\bfr-\bfr'|)\,\frac{g_\rmd(\bfr)g_\rmd(\bfr')}{\Delta_{\rmd\rma}(\bfr)\Delta_{\rmd\rma}(\bfr')}\hat{\Psi}_\rmg^\dagger(\bfr)\hat{\Psi}_\rmg(\bfr)\,\hat{\rho}\,\hat{\Psi}_\rmg^\dagger(\bfr')\hat{\Psi}_\rmg(\bfr')
    \\&=
    \Gamma\frac{\Omega_\rmd^2}{\Delta^2_{\rmd\rma}}\sum_{jk\ell m}K_{jk\ell m}\hat{c}_j^\dagger\hat{c}_k\,\hat{\rho}\,\hat{c}_\ell^\dagger\hat{c}_m,
\end{split}
\end{equation}
where
\begin{equation}\label{eq:spontaneous_tensor}
    K_{jk\ell m} =\int\rmd^2r\rmd^2r'\, j_0(k_0|\bfr-\bfr'|)\,\frac{g_\rmd(\bfr)g_\rmd(\bfr')\varphi_j(\bfr)\varphi_k(\bfr)\varphi_\ell(\bfr')\varphi_m(\bfr')}{[1+ I(\bfr)/\langle I \rangle][1+ I(\bfr')/\langle I \rangle]}.
\end{equation}
From the anticommutator in Eq.~\eqref{eq:spontaneous_emission_general} identical results follow.
At this point two considerations are important.
The dissipator in Eq.~\eqref{eq:spontaneous_dissipator} encodes the entire physics of recoil, heating, and inter-orbital dephasing via the interplay between the weight $g_\rmd(\bfr)/\Delta_{\rmd\rma}(\bfr)$ and the kernel $j_0$.
Second, $K_{jk\ell m}$ is in general a rank-4 tensor, which means that $\mathcal{D}[\hat{L}_\rma]\hat{\rho}$ is not diagonal in creation and annihilation jump operators.
To obtain a diagonal Lindblad form, we can perform a singular value decomposition on $K_{jk\ell m}$ such that
\begin{equation}\label{eq:svd}
    K_{jk\ell m} = \sum_{r=1}^{R}\lambda_\alpha M_{jk}^{(r)}\left[M_{\ell m}^{(r)}\right]^\dagger.
\end{equation}
This allows us to recast Eq.~\eqref{eq:spontaneous_emission_general} as a sum over $R$ jump operators accounting for nonlocal fermionic dephasing,
\begin{equation}\label{eq:dissipator_svd1}
    \mathcal{D}[\hat{L}_\rma]\hat{\rho} = \sum_{r=1}^R\left(\hat{L}_r\,\hat{\rho}\,\hat{L}_r^\dagger - \frac{1}{2}\{\hat{L}_r^\dagger\hat{L}_r,\,\hat{\rho}\}\right),
\end{equation}
with $\hat{L}_r = \sqrt{\Gamma} \frac{\Omega_\rmd}{\Delta_{\rmd\rma}}\sum_{jk}\sqrt{\lambda_r}M_{jk}^{(r)}\hat{c}_j^\dagger\hat{c}_k$, $r=1,\,...,\,\textrm{rank}(K)$.
As we did for the disordered light-matter couplings in Eqs.\eqref{eqs:random_coefficients}, we assume also here in Eq.~\eqref{eq:spontaneous_tensor} that $g_\rmd(\bfr)= g(\bfr)\simeq1$.
Again, we leave for future work a possible optimization of the shape of $g_\rmd(\bfr)$ that decreases the unwanted effects of spontaneous emission.
We remark that the importance of having $\mathcal{D}[\hat{L}_\rma]\hat{\rho}$ in Lindblad form is that we can now easily simulate stochastic quantum trajectories by considering the collection of dephasing jump operators $\hat{L}_\alpha$.
For the experimental parameters we are considering $x_0\ge100\,\textrm{nm}$, i.e., $\eta\sim\mathcal{O}(1)$.
This implies that Eq.~\eqref{eq:spontaneous_tensor} must be used without approximations for an accurate description.
Nevertheless, there are two notable regimes that is worth to discuss.

\subsubsection{1. Deep Lamb-Dicke regime: $\eta\ll1$}

If we fix $\lambda_\rmc$ and thus $k_0$, tuning $x_0$ to zero allows us to reach the deep Lamb-Dicke regime where $\eta\ll1$ and $j_0(k_0|\bfr-\bfr'|)\sim 1$ since $|\bfr-\bfr'|\to0$.
This implies that the spontaneous emission dissipator reduces to
\begin{equation}\label{eq:spontaneous_dissipator_lambd_dicke}
\begin{split}
    \int\rmd^2r\rmd^2r'\hat{L}_\rma(\bfr)\hat{\rho}\hat{L}^\dagger_\rma(\bfr') 
    =
    \Gamma\frac{\Omega_\rmd^2}{\Delta^2_{\rmd\rma}}\sum_{jk\ell m}K_{jk}K_{\ell m}\hat{c}_j^\dagger\hat{c}_k\,\hat{\rho}\,\hat{c}_\ell^\dagger\hat{c}_m,
\end{split}
\end{equation}
where
\begin{equation}\label{eq:lamb_dicke_dissipator}
    K_{jk} = \int\rmd^2r\frac{g_\rmd(\bfr)\varphi_j(\bfr)\varphi_k(\bfr)}{1+ I(\bfr)/\langle I \rangle}.
\end{equation}
The term in Eq.~\eqref{eq:spontaneous_dissipator_lambd_dicke} is qualitatively different from the one in Eq.~\eqref{eq:spontaneous_dissipator}: the photon-coherence kernel is constant (maximal coherence) and the spontaneous emission tensor reduces to a rank-1 tensor that yields a single collective and non-local jump operator $\hat{L}_0 = \sqrt{\Gamma}\frac{\Omega_\rmd}{\Delta_{\rmd\rma}}\sum_{jk}K_{jk}\hat{c}_j^\dagger\hat{c}_k$.
Notice that with spatial uniformity, i.e., no speckle and $g_\rmd(\bfr)=1$, $K_{jk}=\delta_{jk}$ and $\hat{L}_0\propto\sum_j\hat{c}_j^\dagger\hat{c}_j$, so local dephasing only.

\subsubsection{2. Recoil corrections: $\eta\lesssim1$}

To understand the recoil corrections that appear when leaving the deep Lamb-Dicke regime $\eta\ll1$ we consider the $\bfn$-dependent jump operator in Eq.~\eqref{eq:n_jump_operator}.
We perform the adiabatic elimination on $\hat{\Psi}^\dagger_\rmg(\bfr)\hat{\Psi}_\rme(\bfr)$ and expand $\hat{\Psi}_\rmg(\bfr)$ in the harmonic oscillator eigenfunctions.
Since $\eta\lesssim1$, we also expand the exponential in Eq.~\eqref{eq:n_jump_operator} to first order, namely $\rme^{-\rmi k_0\bfn_\perp\cdot\bfr}\simeq1-\rmi k_0\bfn_\perp\cdot\bfr$.
This yields the jump operator
\begin{equation}
    \hat{L}_\rma(\bfn) = \sqrt{\Gamma}\frac{\Omega_\rmd}{\Delta_{\rmd\rma}}\sum_{jk}\left[K_{jk} -\rmi k_0 \int\rmd^2r\frac{g_\rmd(\bfr)(\bfn_\perp\cdot\bfr)}{1+ I(\bfr)/\langle I \rangle} \varphi_j(\bfr)\varphi_k(\bfr)\right]\hat{c}_j^\dagger\hat{c}_k,
\end{equation}
where $K_{jk}$ is given by Eq.~\eqref{eq:lamb_dicke_dissipator}.
Now, $|\bfr|\sim x_0$ over the orbitals.
Therefore, the linear term has magnitude $\sim\eta$ and connects orbitals
differing by one motional quantum, generating incoherent orbital-changing transitions (diffusion/heating) at a rate $\Gamma_{\rm heat}\sim \frac{2}{3}\Gamma_{\rm eff}\eta^2$ with $\Gamma_{\rm eff}=\Gamma(\Omega_\rmd/\Delta_{\rmd\rma})^2$.
The $\frac{2}{3}$ prefactor describes the curvature of the photon-coherence kernel at the origin and hence the mean-square in-plane kick, $\langle
k_\perp^2\rangle/k_0^2=\frac{2}{3}$ for scalar 3D emission.

\subsection{C. Cavity-induced fermionic dephasing}

We derive here Eq.~\eqref{eq:cavity_effective_dissipator} in the main text, following the derivations of Refs.~\cite{uhrich_cavity_2023, baumgartner_quantum_2024, baumgartner_quantum_2025}.
The Schrieffer-Wolff transformation (or, equivalently, the adiabatic elimination on the cavity mode) performed to obtain Eq.~\eqref{eq:hamiltonian_effective} in the main text can be extended to the jump operator of the cavity
\begin{equation}\label{eq:cavity_effective_jump_operator}
    \hat{L}_\rma^{(\rm eff)} = \sqrt{\kappa}\frac{\Omega_\rmd\Omega}{2\Delta_{\rmd\rma}(\Delta_{\rmc\rmd}-\rmi\kappa/2)}\int\rmd^2r\frac{g_\rmd(\bfr)g(\bfr)}{1+ I(\bfr)/\langle I\rangle}\hat{\Psi}_\rmg^\dagger(\bfr)\hat{\Psi}_\rmg(\bfr) = \sqrt{\kappa_{\rm eff}}\sum_{jk}H_{jk}\hat{c}_j^\dagger\hat{c}_k,
\end{equation}
where the effective rate is $\kappa_{\rm eff}=\kappa\frac{\Omega_\rmd^2\Omega^2}{4\Delta_{\rmd\rma}^2(\Delta_{\rmc\rmd}^2 + \kappa^2/4)}\sim\kappa\frac{\Omega_\rmd^2\Omega^2}{\Delta_{\rmd\rma}^2\Delta_{\rmc\rmd}^2}$ and the couplings for the non-local dephasing are given by
\begin{equation}
    H_{jk} = \int \rmd^2 r\,\frac{g_\rmd(\bfr)g(\bfr)\varphi_j(\bfr)\varphi_k(\bfr)}{1+I(\bfr)/\langle I \rangle}.
\end{equation}
The Lindblad dissipator follows straightforwardly (here we write only the first piece, the others are similar)
\begin{equation}
    \hat{L}_\rma^{(\rm eff)}\hat{\rho}\hat{L}^{(\rm eff)\dagger}_\rma = \kappa_{\rm eff}\sum_{jk\ell m}H_{jk}H_{\ell m}\hat{c}_j^\dagger\hat{c}_k\hat{\rho}\,\hat{c}_\ell^\dagger\hat{c}_m,
\end{equation}
that coincides with Eq.~\eqref{eq:cavity_effective_dissipator} in the main text. 
Notice the low rank structure of the above dissipator, compared with the full rank structure of Eq.~\eqref{eq:spontaneous_dissipator}, associated to spontaneous emission away from the Lamb-Dicke regime.

\section{II. COMPARISON BETWEEN f-RQC AND TROTTERIZED PROTOCOLS}

\begin{figure*}[t]
\centering
\includegraphics[width=\textwidth]{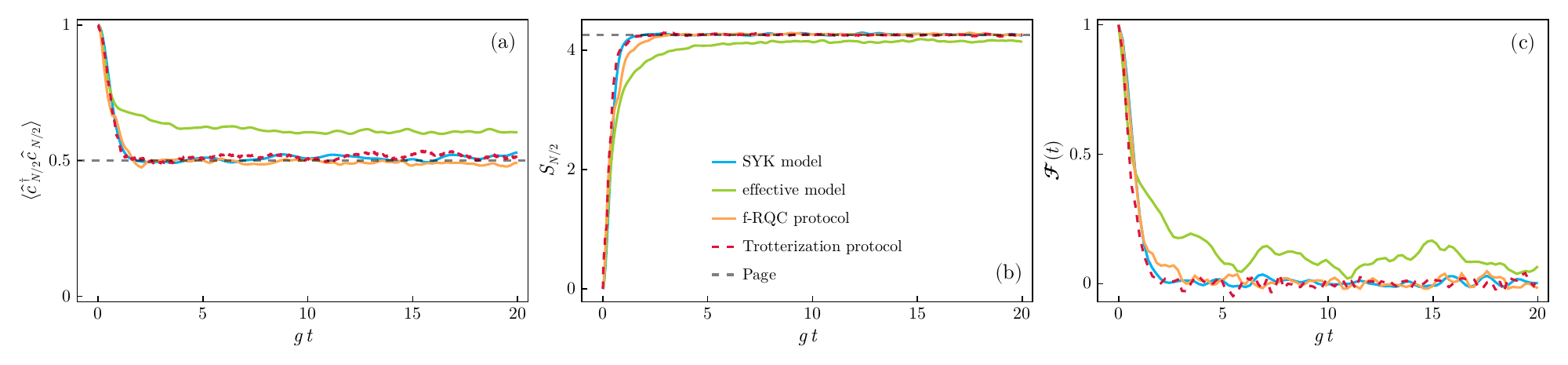}\vspace{0.05em}
\caption{
Comparison among the SYK dynamics generated by the Hamiltonian in Eq.~\eqref{eq:syk} (blue lines) and the dynamics of the effective model in Eq.~\eqref{eq:hamiltonian_effective} with $\Delta_{\rmc\rmd}=1$ and $g_{jk}\sim\mathcal{N}(0, g^2/2N)$ (green lines), the f-RQC protocol described by Eq.~\eqref{eq:f_rqc} (orange lines) and the Trotterization protocol described by Eq.~\eqref{eq:trotterization} (red-dahsed lines). 
The building blocks of both the f-RQC and Trotterization protocols are the effective models with $\Delta_{\rmc\rmd}=1$, while the random couplings are $g_{jk}\sim\mathcal{N}(0, g^2/2N)$ for the f-RQC and $g_{jk}\sim\mathcal{N}(0, g^2/N^2)$ for the Trotterization protocols respectively. 
For the f-RQC simulation, $n=2N+1$, while for the Trotterized simulation, $n_\textrm{T}=100$ and $M=2N$.
We plot (a) the dynamics of the fermionic density $\langle\hat{c}_{N/2}^\dagger\hat{c}_{N/2}\rangle$, (b) the dynamics of the entanglement entropy $S_{N/2}$ and (c) the dynamics of the OTOC in Eq.~\eqref{eq:otoc} with operators $\hat{W}=2\hat{c}_{N/2}^\dagger\hat{c}_{N/2}-1$ and $\hat{V}=2\hat{c}_{N/2+1}^\dagger\hat{c}_{N/2+1}-1$.
The gray-dashed line in panel (b) indicated the Page entanglement entropy for $N$ fermions at half filling.
In all the simulations, we consider $N=14$ fermionic modes at half filling and $g=1$.
}
\label{fig:syk_comparison}
\end{figure*}

In this section we compare the maximally chaotic dynamics generated by the SYK Hamiltonian 
\begin{equation}\label{eq:syk}
     \hat{H}_{\rm SYK}=\sum_{jk\ell m}g_{jk\ell m}\hat{c}_j^\dagger\hat{c}_k^\dagger\hat{c}_\ell\hat{c}_m,
\end{equation}
where $g_{jk\ell m}\sim\mathcal{N}(0, g^2/N^3)$, with the f-RQC time evolution discussed in the main text and the Trotterized time evolution introduced in Ref.~\cite{baumgartner_quantum_2024}, all simulated in the absence of dissipation.
The unitary operator describing the Trotterized time evolution is given by
\begin{equation}\label{eq:trotterization}
    \hat{U}(t) = \textrm{exp}\left[-\rmi\sum_{\alpha=1}^M\hat{H}^{(\alpha)}_{\rm eff} t\right]\simeq \left[\prod_{\alpha=1}^M \rme^{-\rmi\hat{H}^{(\alpha)}_{\rm eff} \Delta t}\right]^{n_\rmT},
\end{equation}
being $\hat{H}^{(\alpha)}_{\rm eff}$ the effective Hamiltonian in Eq.~\eqref{eq:hamiltonian_effective} with a given disorder realization labeled by $\alpha$.
Notably, this implies that if the target model simulation time is $t=n_\rmT\Delta t$, being $n_\rmT$ the number of Trotter steps, its experimental realization requires a physical time equal to $t_{\rm lab} = n_\rmT M\Delta t$.
This needs to be compared to the f-RQC protocol in Eq.~\eqref{eq:f_rqc} for which $t=t_{\rm lab}$.
However, Eq.~\eqref{eq:trotterization} approximates the true SYK dynamics~\cite{baumgartner_quantum_2024, baumgartner_quantum_2025}, while Eq.~\eqref{eq:f_rqc} does not.
The question is then which SYK features are not encoded by the f-RQC protocol that are instead present in the Trotterization protocol by construction.
To answer this question, we consider $N=14$ fermionic modes at half-filling in the absence of dissipation and we study: i) the dynamics of $\langle\hat{c}_j^\dagger\hat{c}_j\rangle$ for $j=N/2$, ii) the entanglement entropy dynamics $S_{N/2}$, and iii) the out-of-time-ordered correlator (OTOC) dynamics
\begin{equation}\label{eq:otoc}
    \mathcal{F}(t) =\langle\hat{W}(t)\hat{V}(0)\hat{W}(t)\hat{V}(0)\rangle,
\end{equation}
where the operators in the OTOC are $\hat{W}=2\hat{c}_{N/2}^\dagger\hat{c}_{N/2}-\mathds{1}$ and $\hat{V}=2\hat{c}_{N/2+1}^\dagger\hat{c}_{N/2+1}-\mathds{1}$ respectively.
As initial state, we consider the product state $\ket{\Psi(0)}=\ket{1\,...\,1\,0\,...\,0}$.

We plot the results in Fig.~\ref{fig:syk_comparison}, where we compare the dynamics generated by Eq.~\eqref{eq:syk} (blue lines) with the effective dynamics generated by Eq~\eqref{eq:hamiltonian_effective} (green lines) where $\Delta_{\rmc\rmd}=1$ and $g_{jk}\sim\mathcal{N}(0, g^2/2N)$, the f-RQC protocol in Eq.~\eqref{eq:f_rqc} (orange lines) and the Trotterization protocol in Eq.~\eqref{eq:trotterization} (red-dashed lines).
For the f-RQC protocol, the random couplings are $g_{jk}\sim\mathcal{N}(0, g^2/2N)$, while for the Trotterization protocol we consider $g_{jk}\sim\mathcal{N}(0, g^2/N^2)$.
In Fig.~\ref{fig:syk_comparison} (a) we plot the dynamics of the fermionic density $\langle\hat{c}_{N/2}^\dagger\hat{c}_{N/2}\rangle$ for these four models.
We observe how the SYK, the f-RQC and the Trotterized dynamics converge to $\langle\hat{c}_{N/2}^\dagger\hat{c}_{N/2}\rangle\sim0.5$ while the effective-model dynamics deviates from the thermal value.
In panel (b) we plot the entanglement entropy $S_{N/2}$.
Again, the SYK, f-RQC and Trotterized models saturate the Page bound $S_{\rm Page} = \frac{N}{2}\log(2) - \frac{1}{2}\log(2) - \frac{1}{4}$, while the effective-model dynamics does not.
Finally, in panel (c) we plot the OTOC dynamics.
We observe how $\mathcal{F}(t)$ rapidly decays to zero for the SYK, f-RQC, and Trotterized dynamics, signaling complete scrambling of quantum information~\cite{swingle_measuring_2016, bohrdt_scrambling_2017, hashimoto_out--time-order_2017, xu_scrambling_2024}.
For the effective model, instead, $\mathcal{F}(t)$ saturates at a finite value.

From these results we conclude that i) the f-RQC model generates many-body quantum chaos but ii) it does not scramble as fast as the SYK or the Trotterized model (notice that the latter yields the SYK time evolution by construction).
If the goal is to control many-body quantum chaos, then the f-RQC eliminates the experimental overhead encoded by the Trotterization approach.
Indeed, having fixed a simulation time $t$, the laboratory time for the f-RQC protocol coincides with $t$, while the laboratory time for Trotterization is $Mt$ with tipically $M\gg1$.
In addition, the Trotterized time evolution is much more exposed to the action of decoherence, especially to heating effects coming from spontaneous emission.

\section{III. PHOTONIC OBSERVABLES}

\begin{figure*}[t]
\centering
\includegraphics[width=0.5\textwidth]{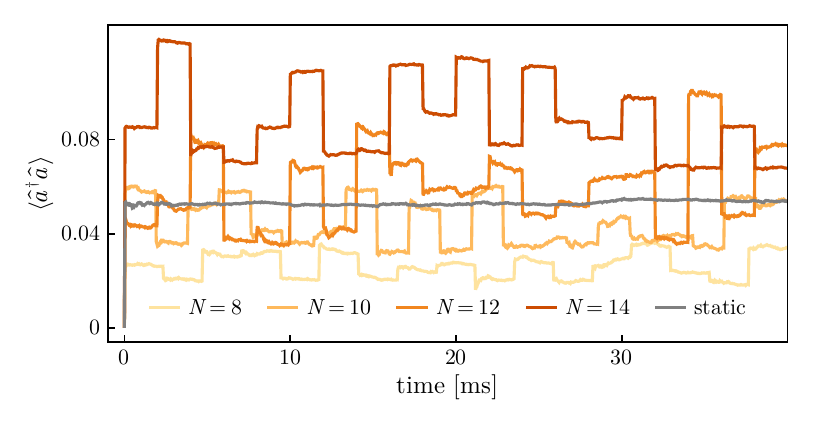}\vspace{0.05em}
\caption{
Time evolution of the cavity photon number $\langle\hat{a}^\dagger\hat{a}\rangle$ with the f-RQC protocol for $N=8, 10, 12, 14$ fermions (from light to dark orange lines) and in the presence of static disorder for $N=12$ (gray line).
The average is taken over $N_{\rm traj}=100$ trajectories. 
All parameters as in Fig.~\ref{fig:small_detuning}.
}
\label{fig:photonic_observables}
\end{figure*}

\begin{figure*}[t]
\centering
\includegraphics[width=0.4\textwidth]{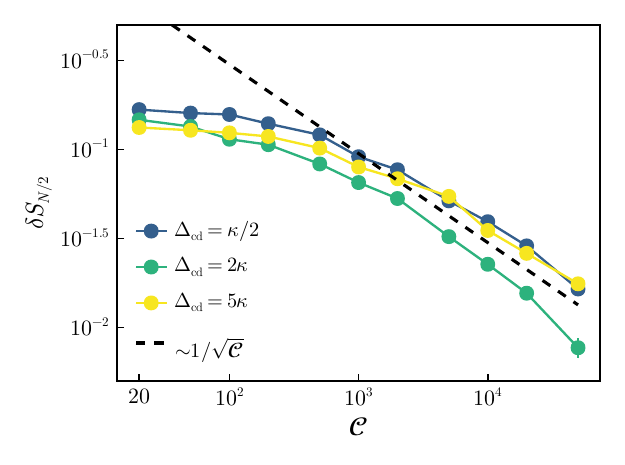}\vspace{0.05em}
\caption{
Relative distance from Page entropy $\delta S_{N/2} = |S_{N/2}-S_{\rm Page}|/S_{\rm Page}$ as a function of $\mathcal{C}$ for $\Delta_{\rm cd}=\kappa/2$ ($\Omega_\rmd/2\pi=12\,\textrm{MHz}$), $\Delta_{\rm cd}=2\kappa$ ($\Omega_\rmd/2\pi=25\,\textrm{MHz}$) and $\Delta_{\rm cd}=5\kappa$ ($\Omega_\rmd/2\pi=38\,\textrm{MHz}$) at fixed $N=10$.
Data are averaged over $N_{\rm traj}=100$ trajectories and 5 disorder realizations.
The black-dashed line represents the asymptotic behavior $\sim1/\sqrt{\mathcal{C}}$.
Other parameters as in Fig.~\ref{fig:small_detuning}.
}
\label{fig:ee_detuning}
\end{figure*}

Our treatment also gives access to photonic observables, including the cavity emission, one of the simplest experimental probes in cavity QED.
In Fig.~\ref{fig:photonic_observables} we plot the time evolution of $\langle\hat{a}^\dagger\hat{a}\rangle$ corresponding to the system configuration of Fig.~\ref{fig:small_detuning} and take the average over $N_{\rm traj}=100$ Monte Carlo quantum trajectories.
The gray lines corresponds to the time evolution with static disorder, as in  Fig.~\ref{fig:small_detuning} (a).
The cavity field rapidly reaches the steady state at a value $\langle\hat{a}^\dagger\hat{a}\rangle\ll1$, thus confirming the validity of the dispersive approximation.
The orange lines correspond instead to the f-RQC time evolution for $N=8, 10, 12, 14$ fermions (from light to dark).
The cavity field responds to the instantaneous change of the light-matter couplings $g_{jk}$ with a sudden shift in its population.
For all the cases considered here, $\langle\hat{a}^\dagger\hat{a}\rangle\ll1$ also for the f-RQC protocol.
We observe also that, as expected, the cavity photon number grows on average with $N$.

\section{IV. ENTANGLEMENT SCALING AT VARIOUS $\Delta_{\rm cd}$}

In this section we complement the analysis on the entanglement entropy scaling with the cavity cooperativity $\mathcal{C}$ presented in the main text.
In Fig.~\ref{fig:ee_detuning} we plot the relative distance from the Page entanglement entropy $\delta S_{N/2} = |S_{N/2}-S_{\rm Page}|/S_{\rm Page}$ for $N=10$ fermionic modes at half filling for different values of the cavity-drive detuning $\Delta_{\rm cd}$.
We adjust the drive amplitude $\Omega_\rmd$ to keep the effective coherent energy scale $\mathcal{E}$ constant.
We observe how the scaling $1/\sqrt{\mathcal{C}}$ is recovered for $\mathcal{C}\gtrsim10^3$ regardless of the value of $\Delta_{\rm cd}$ (encoding which dissipation source is dominating).
For realistic values $\mathcal{C}\lesssim100$ the scaling of $\delta S_{N/2}$ with $\mathcal{C}$ is much slower than $1/\sqrt{\mathcal{C}}$ and the relative error with the Page value stays well above $10\%$, confirming that cooperativities of few tens are not sufficient to generate massively entangled states, at least for the protocol we considered throughout this article.

\section{V. MANY-BODY DYNAMICS WITHOUT ATOMIC SPONTANEOUS EMISSION}

\begin{figure*}[t]
\centering
\includegraphics[width=\textwidth]{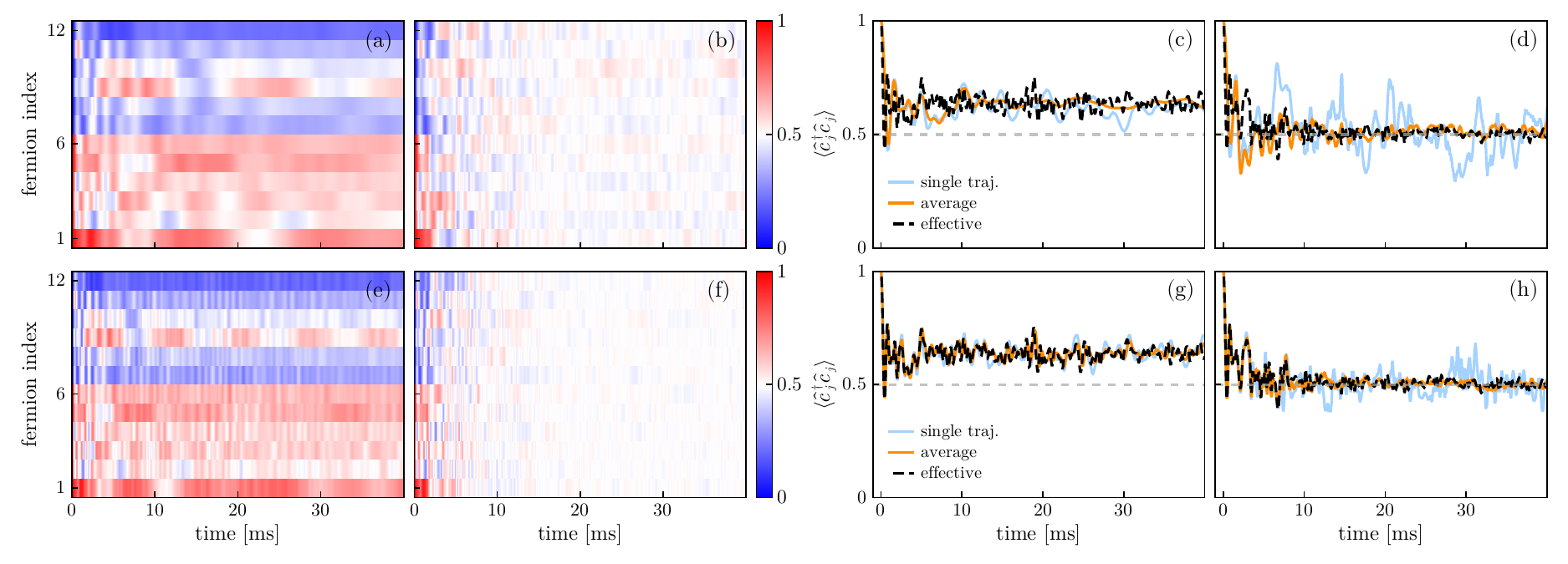}\vspace{0.05em}
\caption{
Quantum many-body dynamics with cavity dissipation as the sole decoherence channel for $N=12$ fermionic modes at half filling.
(a-d) Dynamics for $\Delta_{\rmc\rmd}=\kappa/2$.
(e-h) Dynamics for $\Delta_{\rmc\rmd}=5\kappa$.
Panels (a) and (e) display the time evolution of $\langle\hat{c}_j^\dagger\hat{c}_j\rangle$ for static disorder.
Panels (b) and (f) display the time evolution of $\langle\hat{c}_j^\dagger\hat{c}_j\rangle$ for the f-RQC protocol.
Panels (c), (d), (g) and (h) compare the dissipative quantum trajectory averaged dynamics (orange curves), the single-trajectory averaged dynamics (light blue curves) and the effective dynamics described by Eq.~\eqref{eq:hamiltonian_effective} with $\kappa=0$ (black-dahsed curves). 
Panels (c) and (g) refer to static disorder while (d) and (h) to the f-RQC protocol.
The curves are for $j=6$.
Other parameters are fixed as in Fig.~\ref{fig:small_detuning}.
}
\label{fig:cavity_dissipation}
\end{figure*}

To further illustrate the harmful role played by atomic spontaneous emission, we plot here the quantum many-body dynamics with cavity dissipation as the sole decoherence source for $N=12$ fermions at half filling for $\Delta_{\rm cd}=\kappa/2$ and $\Delta_{\rm cd}=5\kappa$ (the two detuning values considered in the main text) and $x_0=300\,\textrm{nm}$.

In Fig.~\ref{fig:cavity_dissipation} (a-d) we plot the results for $\Delta_{\rmc\rmd}=\kappa/2$.
Figs.~\ref{fig:cavity_dissipation} (a) and (b) display the dynamics of $\langle\hat{c}_j^\dagger\hat{c}_j\rangle$ with static disorder and the f-RQC protocol respectively. 
The distinction between integrable and chaotic dynamics via thermalization of local observables appears more distinct than its counterpart in Fig.~\ref{fig:small_detuning} that includes atomic spontaneous emission, even if the latter is suppressed by the small value of $\Delta_{\rmc\rmd}$.
When comparing the effective-model dynamics with the dissipative dynamics for a single quantum trajectory or averaged over $N_{\rm traj}=100$ trajectories, we observe how the open system dynamics is slowed down with respect to the Hamiltonian time evolution.
In this regime, the cavity is frequently monitoring the fermionic system, effectively acting with the single, sparse jump operator in Eq.~\eqref{eq:cavity_effective_jump_operator}, in a way that is reminiscent of the quantum Zeno effect~\cite{facchi_quantum_2008}.
As already stated in the main text, the cavity monitoring does not qualitatively change the behavior of the fermionic density, but rather disentangles the otherwise maximally entangled unitary dynamics

In Fig.~\ref{fig:cavity_dissipation} (e-h) we plot the results for $\Delta_{\rmc\rmd}=5\kappa$.
Compared to the case with $\Delta_{\rmc\rmd}=\kappa/2$, the distinction between chaos and integrability is even more evident and, importantly, the averaged open system dynamics coincides with the effective unitary dynamics, as one can see in Figs.~\ref{fig:cavity_dissipation} (g) and (h).
This is a direct consequence of the suppression of $\kappa_{\rm eff}$ with an increased $\Delta_{\rmc\rmd}$.
The dynamics over single quantum trajectories still deviates from the unitary prediction, even though oscillations are suppressed with respect to the case with smaller detuning.
Only when $\Delta_{\rmc\rmd}\gg\kappa$ single quantum trajectories coincide with the effective model dynamics, and the Page entanglement bound is saturated by the dissipative dynamics as well (not shown).
In this regime, the driven-dissipative cavity QED setup is able to maximally entangle the fermionic particles.
Unfortunately, this coincides with the regime that is dominated by atomic spontaneous emission, and the only parameter that can open a genuine window for the quantum simulation of maximally entangled fermionic systems in cavities is the cooperativity $\mathcal{C}$.
$\mathcal{C}=20$ is however nevertheless sufficient for the experimental observation of simple fingerprints of the phenomenon.

\section{VI. ADDITIONAL DETAILS ON THE SPONTANEOUS EMISSION TREATMENT}

\begin{figure*}[t]
\centering
\includegraphics[width=0.8\textwidth]{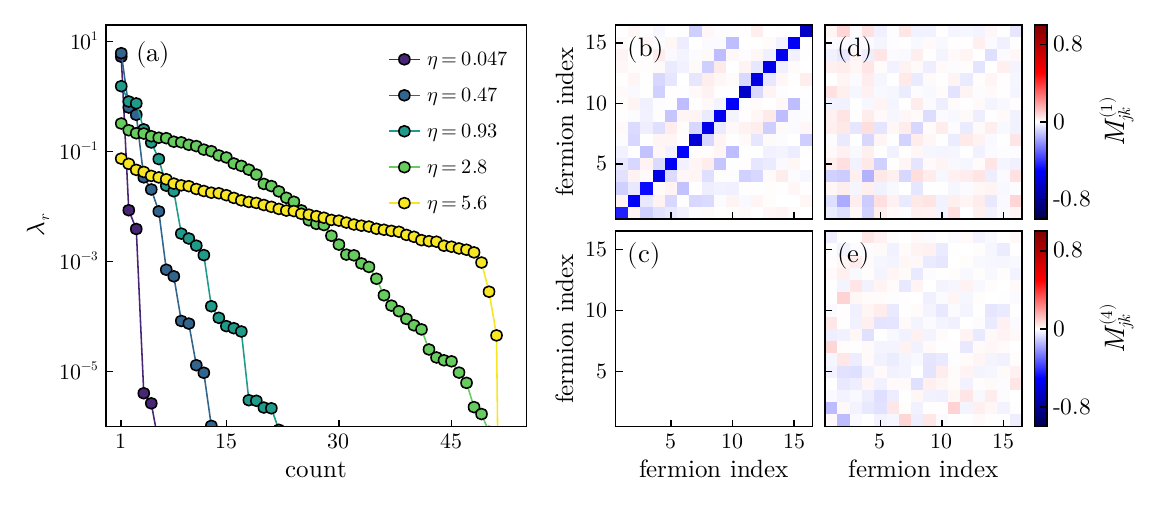}\vspace{0.05em}
\caption{
Behavior of the spontaneous emission couplings with the Lamb-Dicke parameter $\eta$.
(a) Singular values $\lambda_r$ coming from the decomposition of the spontaneous emission tensor $K_{jk\ell m}$ as in Eqs.~\eqref{eq:spontaneous_tensor} and \eqref{eq:svd} for several values of $\eta$ from $\eta\ll1$ (deep Lamb-Dicke regime) to $\eta\sim\mathcal{O}(5)$.
(b-e) Couplings $M_{jk}^{(r)}$ for (b) $\eta=0.047$ and $r=1$, (c) $\eta=0.047$ and $r=4$, (d) $\eta=2.8$ and $r=4$, (e) $\eta=2.8$ and $r=4$.
The number of fermionic modes is fixed to $N=16$.
}
\label{fig:dissipative_coefficients}
\end{figure*}

In this section, we study more in detail the crossover from the deep Lamb-Dicke regime $\eta\ll1$, experimentally unachievable in typical cavity QED experiments, and the regime where $\eta\sim{O}(1)$, which is instead a typical operating point of these platforms.
In Fig.~\ref{fig:dissipative_coefficients} (a) we study the behavior of the singular values obtained after the decomposition of the spontaneous emission tensor $K_{jk\ell m}$ as in Eqs.~\eqref{eq:spontaneous_tensor} and \eqref{eq:svd} for various $\eta=k_0x_0$ where $k_0=2\pi/\lambda_\rmc$ with $\lambda_\rmc=671\,\textrm{nm}$.
When $\eta\ll1$ (the deep Lamb-Dicke regime) we obtain a single dominant eigenvalue $\lambda_0$ which yields a single non-local dephasing jump operator $\hat{L}_0$. 
The other singular values immediately drop below $10^{-2}$ and rapidly reach zero within numerical precision.
As $\eta$ becomes larger, more decoherence channels are opened and the number of jump operators needed to describe spontaneous emission increases (namely, the rank of the tensor $K_{jk\ell m}$ grows).
When $\eta\gtrsim1$, a large number of almost equivalent dephasing jump operators participate in the dissipative dynamics.
This feature ultimately encodes the harmful character of atomic spontaneous emission on the many-body quantum dynamics we studied in the main text.
Notably, our treatment supposes a truncated orbital basis, and we refer the Reader to the End Matter for a discussion on the importance of trap engineering to avoid such effects in the actual experiment. 

Finally, in Figs.~\ref{fig:dissipative_coefficients} (b-e), we plot the spontaneous emission coefficients $M_{jk}^{(r)}$ for $\eta=0.047$ [panels (b) for $r=1$ and (c) for $r=4$] and $\eta=2.8$ [panels (d) for $r=1$ and (e) for $r=4$].
We observe how in the deep Lamb-Dicke regime we have a single dominant jump operator with a large diagonal part.
Here, off-diagonal elements are due to the spatially inhomogeneous speckle pattern.
When $\eta\gtrsim1$, a much larger number of jump operators emerges.
Figs.~\ref{fig:dissipative_coefficients} (d) and (e) show how $M_{jk}^{(r)}$ describe non-local dephasing processes with the same strength for both $r=1$ and $r=4$, as well as larger values of $r$ (not shown).

\end{document}